\documentclass[11pt,a4paper]{article}
\pdfoutput=1
\usepackage{jheppubnohead}
\usepackage[utf8]{inputenc}
\usepackage{amsmath}
\usepackage{epsfig}
\usepackage{graphicx}
\usepackage{amssymb}
\usepackage{tabularx}
\usepackage[normalem]{ulem} 
\usepackage{booktabs} 
 
\def\beqn{\begin{eqnarray}}
\def\eeqn{\end{eqnarray}} 
\def\be{\begin{equation}}
\def\ee{\end{equation}}
\def\nn{\nonumber}

\newcommand{\lbo}{\lambda_{B_1}} 
\newcommand{\lbt}{\lambda_{B_2}} 
\newcommand{\lxo}{\lambda_{\chi_1}} 
\newcommand{\lxt}{\lambda_{\chi_2}} 
\newcommand{\lp}{\lambda_\Psi} 
\newcommand{\mx}{m_\chi} 
\newcommand{\mf}{m_U} 
\newcommand{\mq}{m_Q} 
\newcommand{\mps}{m_\Psi} 
\newcommand{\mso}{m_{S_1}} 
\newcommand{\mst}{m_{S_2}} 
\newcommand{\mh}{m_h}

\providecommand{\proarrow}[0]{\rightarrow}
\providecommand{\dif}[0]{\mathrm{d}}
\providecommand{\proname}[2]{#1 \proarrow #2}
\providecommand{\lrproname}[2]{#1 \leftrightarrow #2}
\providecommand{\order}[1]{O\left( #1 \right)}
\providecommand{\mibr}[2]{\text{Br}(\proname{#1}{#2})}
\providecommand{\ghor}[2]{\gamma \left(\proname{#1}{#2}\right)}
\providecommand{\gphor}[2]{\gamma' \left(\proname{#1}{#2}\right)}
\providecommand{\dghor}[2]{\Delta \gamma \left(\proname{#1}{#2}\right)}
\providecommand{\dgphor}[2]{\Delta \gamma' \left(\proname{#1}{#2}\right)}
\providecommand{\ydyeq}[1]{\frac{Y_{#1}}{Y^{eq}_{#1}}}
\providecommand{\ydyeqs}[1]{\frac{Y_{#1}^2}{Y^{eq\, 2}_{#1}}}
\providecommand{\myexp}[1]{e^{#1}}

 \title{On baryogenesis from dark matter annihilation}
 
 \author[a]{Nicol\'as~Bernal,}
 \author[a,b]{Stefano~Colucci,}
 \author[c]{Fran\c{c}ois-Xavier~Josse-Michaux,}
 \author[d]{J. Racker,}
 \author[a]{Lorenzo~Ubaldi}

 \affiliation[a]{Bethe Center for Theoretical Physics and Physikalisches Institut, \\
Universit\"at Bonn, Nu\ss allee 12, D-53115 Bonn, Germany}
 \affiliation[b]{Department of Physics, University of Turin, via P. Giuria 1, 10125 Turin, Italy}
 \affiliation[c]{Centro de F\'isica Te\'orica de Part\'iculas CFTP, Instituto Superior T\'ecnico, \\
Technical University of Lisbon, 1049-001 Lisbon, Portugal}
\affiliation[d]{Instituto de F\'isica corpuscular (IFIC), Universidad de
Valencia-CSIC \\ 
Edificio de Institutos de Paterna, Apt. 22085, 46071 Valencia,
Spain}

\emailAdd{nicolas@th.physik.uni-bonn.de}
\emailAdd{colucci@th.physik.uni-bonn.de}
\emailAdd{fxjossemichaux@gmail.com}
\emailAdd{racker@ific.uv.es}
\emailAdd{ubaldi@th.physik.uni-bonn.de}

\abstract{We study in detail the conditions to generate the baryon asymmetry of the universe from the annihilation of dark matter. This scenario requires a low energy mechanism for thermal baryogenesis, hence we first discuss some of these mechanisms together with the specific constraints due to the connection with the dark matter sector. Then we show that, contrary to what stated in previous studies, it is possible to generate the cosmological asymmetry without adding a light sterile dark sector, both in models with violation and with conservation of $B-L$. In addition, one of the models we propose yields some connection to neutrino masses.
}

\begin{document}
\hfill {\tt Bonn-TH-2013-11}

\hfill {\tt CETUP2013-010}

\hfill {\tt CFTP/13-015}

\hfill {\tt IFIC/13-49}

\maketitle

\section{Introduction}

We have evidence that most of the matter content of the universe is in a non-luminous form called Dark Matter (DM). The remaining, less abundant, visible matter is well understood at the fundamental level in the theoretical framework of the Standard Model (SM) of particle physics. However the fact that in the latter sector we observe an excess of matter over anti-matter, the Baryon Asymmetry of the Universe (BAU), cannot be explained within the SM. It is clear that both DM and BAU require physics Beyond the Standard Model (BSM) and there are many ideas as to what the DM candidate could be and what mechanism could be responsible for the BAU.

In principle the DM and the BAU problems could be unrelated. Indeed they have often been approached separately in the literature. Nevertheless one could entertain the idea that they have a common solution that can be found within the same model. One such possibility is in the framework of Asymmetric Dark Matter (ADM)~\cite{Nussinov:1985xr, Roulet:1988wx, Barr:1990ca}, where one speculates that in the dark sector there is a matter anti-matter asymmetry that is related to the one in the visible sector, the BAU. This connection is motivated by the fact that the ratio of the abundances of dark and baryonic matter, $\Omega_{\rm DM} / \Omega_{\rm B} \sim 5$, is a number not far from one, which would suggest a common mechanism for the origin of the two species. 

A different possibility for the DM candidate is provided by a Weakly Interacting Massive Particle (WIMP). Generically, a WIMP annihilates with weak-force strength into SM particles and it is in thermal equilibrium in the early hot universe. Eventually, as the universe expands, these annihilations freeze out and the DM abundance is set. The WIMP is referred to as a thermal relic. It is a striking coincidence that if a massive particle has interactions at the electroweak scale, this results automatically in the thermal relic abundance that matches the observed value. Such a coincidence is dubbed the ``WIMP miracle''. The simplicity of this argument makes the WIMP a very well motivated and the most studied DM candidate.

One can ask the following questions. Can we generate the BAU through WIMP annihilations? If so, what are the conditions for it to be possible? The first question has already been positively answered in previous work~\cite{Cui:2011ab, Bernal:2012gv}, where the WIMPy baryogenesis mechanism has been proposed. In this paper we will critically address the second one to understand which of the conditions described in Refs.~\cite{Cui:2011ab, Bernal:2012gv} are crucial and which can be relaxed.

Before turning to it, we want to emphasize that WIMPy baryogenesis does not predict that the ratio of abundances $\Omega_{\rm DM} / \Omega_{\rm B}$ is of order one. Instead, it was estimated in Ref.~\cite{Cui:2011ab} that one can end up in the following range
\be
10^{-1} < \frac{\Omega_{\rm DM}}{\Omega_{\rm B}} < 10^6\nonumber,
\ee
so that the observed value can be accommodated. The motivation for exploring this mechanism is rather given by the fact that the simple WIMP paradigm can be easily extended in order to generate the BAU. Since the baryon asymmetry is typically generated at temperatures below a few TeV, this provides an interesting example of low-energy thermal baryogenesis.

In this regard, it is worth noting that in most ADM models, although the origin of the baryonic and DM sectors are closely interrelated, the ratio of the corresponding abundances is also not really predicted. These models, in their simplest version, predict that the number densities of DM and baryons are similar, $n_{\rm DM} \simeq n_B$. Then, to get $\Omega_{\rm DM} / \Omega_{\rm B} \sim 5$, one needs a DM particle with a mass of order 5 GeV, i.e. of the same order of the proton mass. But the proton mass is determined by QCD interactions, so there is a priori no reason at all for the DM to have this mass (for some exceptions see e.g.~\cite{An:2009vq,Blennow:2010qp,Bai:2013xga}). In other words, the puzzling similarity among $\Omega_{\rm DM}$ and $\Omega_{\rm B}$ has an explanation in terms of another miraculous relation, namely $m_{\rm DM} \sim m_{\rm proton}$. In turn, ADM models with heavier DM particles rely on a Boltzmann suppression of the DM asymmetry, so again the ``factor 5'' is not explained, or to put it another way, many different values of $\Omega_{\rm DM} / \Omega_{\rm B}$ could be obtained. Furthermore, there are severe constraints from direct detection experiments~\cite{MarchRussell:2012hi}, due to the fact that ADM must have larger interactions than ordinary WIMPs in order to efficiently annihilate away the symmetric component. These issues for ADM models provide further motivation for exploring other mechanisms that can yield a relation between the DM and BAU, like those proposed in Refs.~\cite{McDonald:2011sv, McDonald:2011zza, D'Eramo:2011ec, Canetti:2012vf, Cui:2011ab, Davidson:2012fn}.

The paper is organized as follows. In Section~\ref{sec:conditions} we briefly review the ingredients of the WIMPy baryogenesis models explored so far, we critically examine what conditions are crucial for the success of the mechanism, we point out that some assumptions made in previous work can actually be relaxed, and we claim that there are some interesting variants still to be explored in the context of Baryogenesis from Dark Matter Annihilation (BarDaMA). In Sections~\ref{sec:model} and~\ref{sec:modelC} we present models that show explicitly how such variants work. We then conclude and include some appendices with technical details.


\section{Conditions for BarDaMA } \label{sec:conditions}

Throughout this paper by WIMP we denote a SM gauge singlet, weakly interacting massive particle that gives the right DM abundance as a thermal relic. In this section we first give a brief overview of the WIMPy baryogenesis mechanism as proposed in Ref.~\cite{Cui:2011ab}, then we ask generally what conditions are needed and what assumptions can be relaxed.

The mechanism we are interested in was first proposed in Ref.~\cite{Cui:2011ab} and further studied with an effective operator analysis in Ref.~\cite{Bernal:2012gv}. In a nutshell: two WIMPs, that we denote by $\chi$, annihilate into a SM quark and an exotic, heavy antiquark, $\Psi$. The latter then decays into two SM quarks and a light ($<$ eV) sterile SM singlet fermion, $n$. This decay can either be explicitly $B$-violating (if $n$ has zero baryon number, $B$) or $B$-conserving. In that case $n$ has to carry some $B$ charge, which is then sequestered into an invisible dark sector and results in an effective $B$ violation in the visible sector. $CP$ violation is provided in the WIMP annihilation via a combination of complex couplings and interference between tree-level and one-loop diagrams, as shown for example in Fig.~\ref{Fig:CPandwashout}. Departure from thermal equilibrium is guaranteed by being in the proximity of the WIMP freeze-out. Thus, the three Sakharov conditions are fulfilled~\cite{Sakharov:1967dj} and a baryon asymmetry is generated. Quantitatively, the  BAU yield is tied to usual washout effects and to the decay of the exotic $\Psi$. The one just described is the simplest version of the mechanism. A similar leptogenesis variant was also presented in Ref.~\cite{Cui:2011ab}.

The scale for baryogenesis in these models is given by the WIMP mass, $m_\chi$, since the asymmetry starts to be produced at $T \lesssim m_\chi$, when the WIMP becomes non-relativistic.
The lower bound on $m_\chi$ depends on whether the WIMP annihilates into leptons or into quarks. In the former case $m_\chi$ has to be greater than $\order{1}$ TeV in order to allow sphalerons to convert lepton number into baryon number~\cite{Cui:2011ab}. In the latter case, lower masses are possible, but one has to confront with LHC bounds on the new colored particle, $\Psi$, produced in the annihilation. This implies a lower limit\footnote{The limit quoted in Ref.~\cite{Bernal:2012gv} was of 400 GeV. Updated limits from ATLAS~\cite{ATLAS-CONF-2012-109} result in a bound of $\sim 1$ TeV on $m_\Psi$, which in turn translates into a bound of 500 GeV on $m_\chi$. These constraints apply to the models of Refs.~\cite{Cui:2011ab, Bernal:2012gv}, but do not apply to the models we are going to study in this work, as we explain in Section~\ref{sec:model}.} of 500 GeV on $m_\chi$.

In addition, unitarity arguments~\cite{Griest:1989wd} set an upper limit on $m_\chi$ of 340 TeV, which is a pretty low energy scale for thermal baryogenesis mechanisms.
In fact, it is known that having thermal baryogenesis from particle decays or annihilations is challenging at these low temperatures, $T \lesssim 100$~TeV. The main problem stems from the fact that CP violation requires not only a complex phase in the couplings, but also a kinematical phase. This one in turn implies the existence of on-shell processes that violate $L$ or $B$, as shown for example in Fig.~\ref{Fig:CPandwashout}, and tend to washout the $B-L$ asymmetry. Since the CP asymmetry is proportional to the couplings of this $L$/$B$ - violating interactions, they cannot be very small, which leads to washout processes that are typically too fast compared to the expansion rate of the universe if baryogenesis occurs at low temperatures.

There are some ways to get around the difficulties of low-energy thermal baryogenesis, which we discuss in connection to WIMP DM in the next subsection. We will find that the most attractive possibility is to include a massive field in the annihilation products, and we will go on to examine the corresponding model building conditions in subsection~\ref{sec:massiveann}.

\subsection{Exploring ways for BarDaMA}
We describe here four possibilities to achieve thermal baryogenesis from particle decays or annihilations around the TeV scale~\cite{Hambye:2001eu, Cui:2011ab, Racker:2013lua}. As we explain next, the connection to the DM sector imposes serious restrictions on some of them, leaving the last one as the most attractive.
\begin{itemize}
\item[(I)] For temperatures above $\sim 10^2$~TeV it is possible to generate the BAU from CP-violating annihilations of heavy particles into SM particles, without resorting to any of the mechanisms described below. However, for this to happen the couplings of baryons (or leptons) to the mediator of the annihilation cannot be very large. We verified that the resulting DM relic density would be several orders of magnitude above the observed one. Therefore we do not pursue this possibility any further. 

\item[(II)] In~\cite{Hambye:2001eu} it was proposed to generate the BAU in the three-body decay of a heavy particle. The basic idea is that washout processes involving three particles in the initial or final state are naturally phase-space suppressed with respect to $1 \leftrightarrow 2$ and $2 \leftrightarrow 2$ interactions, while the CP asymmetry could be still sizeable. On one hand, it would be interesting to confirm that actually all washout processes can be suppressed without reducing too much the CP asymmetry; on the other hand, it seems difficult to extend this mechanism to BarDaMA, given that $2 \to 3$ annihilations would be suppressed, presumably yielding a too large relic DM density.  

\item[(III)] When the CP asymmetry is induced by a pair of particles almost degenerate in mass, it can be enhanced up to $\order{1}$ values~\cite{flanz96, covi96II, pilaftsis97II}. This resonant mechanism has been widely studied for baryogenesis from heavy particle decays, especially in leptogenesis models (see e.g. Ref.~\cite{pilaftsis03}). The washouts are suppressed simply by taking the relevant couplings small enough, while the CP asymmetry is kept large due to the mass degeneracy of the heavy particles. It is interesting to wonder if this mechanism can also work for BarDaMA. In doing so, we find two difficulties: \\
(a) Not only the two mediators ($S_1$ and $S_2$) of DM annihilations have to be almost degenerate in mass, but the DM mass, $m_{\chi}$, must also satisfy  $ 2\,m_{\chi} \sim m_{S1}$. This last condition comes from setting $s\sim m_{S_1}^2$ in the scalar propagator, which is necessary to get the enhancement, and from the fact that for non-relativistic DM we have $\sqrt{s} \simeq 2\,m_\chi + \order{T}$, where the temperature $T$ is much smaller than $m_\chi$ near freeze-out. \\
(b) More importantly, suppressing the washouts by taking the couplings between ${S_{1,2}}$ and the baryons (or leptons) small enough, also reduces the DM annihilation rate, which may yield too much relic DM. We have made some rough quantitative estimates and found that indeed it is very challenging to get the correct relic DM density while suppressing the washouts this way. \\
It is out of the scope of this work to perform precise -and subtle- calculations to establish whether or not the resonant mechanism can yield a successful  BarDaMA.  Its main interest lies in that it may open some alternatives to those studied here and in previous works.
\item[(IV)] Another way to obtain the BAU at low energies is to include a massive field, $\Psi$, in the annihilation products, with $m_{\Psi} \gtrsim m_{\chi}$~\cite{Cui:2011ab}. Then the washouts can be Boltzmann suppressed, while the CP asymmetry can be sizeable as long as  $m_{\Psi}$ is not too close to $2\,m_{\chi}$. Some concrete models have been presented in~\cite{Cui:2011ab}, with DM annihilation producing $\Psi$ + SM quarks or $\Psi$ + SM leptons. In either case $\Psi$ cannot be a SM singlet. To avoid gauge anomalies, one takes $\Psi$ to be vector-like, that has also the advantage of allowing a large mass term, $m_\Psi$, not related to the electroweak symmetry breaking. When $m_{\chi} \lesssim  m_{\Psi} < 2\,m_{\chi}$ this mechanism is simple and efficient\footnote{Here the upper bound, $m_\Psi < 2\,m_\chi$ is dictated by kinematics and is strict. On the contrary, the lower bound can be somewhat relaxed, as some of us pointed out in Ref.~\cite{Bernal:2012gv}. Values down to $m_\Psi \sim 0.5 \ m_\chi$ can still permit to achieve the observed BAU and DM density in some models. } . 
\end{itemize}

\subsection{BarDaMA with massive annihilation products}\label{sec:massiveann}
Motivated by the above discussion, we now only focus on option (IV), in which the DM annihilation products contain an exotic heavy field, $\Psi$, that leads to Boltzmann suppressed washouts. In order to avoid overclosing the universe $\Psi$ has to decay. It turns out that it is far from trivial to make $\Psi$ decay without erasing the baryon asymmetry produced in the annihilations, and due to this requirement WIMPy baryogenesis models become more involved. To address this issue in detail, we find it convenient first to cast the baryon and/or lepton number violation in terms of the quantity $B-L$, that is conserved by non-perturbative sphaleron processes~\footnote{Note that a violation of only $B$ or $L$ corresponds to a violation of $B-L$ as well.}. Second we separate the discussion into two mutually exclusive cases: in the Lagrangian, (a) there are no operators that violate $B$ and/or $L$, and (b) there are such operators.
\begin{itemize}
\item[(a)] {\it B-L conserving case}: In these scenarios it is possible to assign a $B$ and an $L$ charge to $\Psi$, $B_\Psi$ and $L_{\Psi}$, so that there exists a total conserved $B-L = (B-L)_{SM} + (B-L)_{\Psi} + (B-L)_{\text{other fields}}$, where $(B-L)_{SM}$ stands for the sum of the $B-L$ charges of all SM particles and $(B-L)_{\text{other fields}}$ for that of other fields that may exist (and actually have to exist as we will explain in the next section). Since we want to explain dynamically the origin of the BAU, we take $Y_{B-L}=0$ as an initial condition and therefore $Y_{B-L}$ remains always null. Here $Y$ is the usual ratio of number density over entropy density, $Y \equiv n/s$.

In Ref.~\cite{Cui:2011ab} the authors argued that $\Psi$ should decay into light sterile particles, decoupled from the SM fields at low temperatures, otherwise the $(B-L)_{SM}$ asymmetry would be erased. This way the universe would contain  today  a matter-antimatter asymmetry in the SM fields, and another one of opposite sign in a ``sequestered'' sector. To avoid decays of $\Psi$ entirely into SM particles, that could erase the asymmetry, they imposed a discrete symmetry. Together with the requirement of DM stability, the discrete symmetry had to be at least a $\mathbb{Z}_4$~\cite{Cui:2011ab, Bernal:2012gv}, in contrast with many simple DM models which typically require just a $\mathbb{Z}_2$. 

Since we want to determine the conditions for BarDaMA, we address the question of whether or not the requirements mentioned above are necessary. We find that there is actually a loophole in the argument, which allows for WIMPy baryogenesis with a conserved $B-L$ and no sequestered sector. The key is provided by the freeze-out of the sphalerons, which occurs at temperatures close to the critical temperature of the electroweak phase transition~\cite{Kuzmin:1985mm,Burnier:2005hp}. To see the essential idea consider a model where $L=L_{SM}$ is only violated by sphaleron processes  (i.e. $L_{SM}$ is conserved at the perturbative level).
Although $B-L$ is always null, some asymmetry can be generated in baryons during DM annihilations, in which case this gets rapidly redistributed over the fields that are -almost- in thermal equilibrium. Notably, fast sphaleron processes transform part of the $B_{SM}$ asymmetry into a $L_{SM}$ one. If sphalerons freeze out while there is some asymmetry in the exotic fields, $L_{SM}$ will remain constant and generally not null until the present time. Moreover, the only {\em stable} particles that carry $B-L$ charge are chosen to belong to the SM. Therefore once all the exotic heavy particles have decayed, $0 = B-L = B_{SM}-L_{SM}$, i.e. $B_{SM}=L_{SM}$, and consequently it is possible to generate a cosmological baryon asymmetry without a sequestered sector (instead the conservation laws lead to a universe with an equal amount of lepton and baryon asymmetry). For the final baryon asymmetry to be significant, it is crucial that most of the exotic particles decay after sphalerons freeze out. We remark that an analogous mechanism holds interchanging the roles of leptons and baryons. Also it is worth noting that this same effect has been used to generate the BAU from singlet neutrino decays in leptogenesis models with conserved $B-L$~\cite{GonzalezGarcia:2009qd}.

Actually, as we explain next, it is not so straightforward to generate the BAU with the mechanism just described. Nevertheless it can be done, as we show with a concrete model in Sec.~\ref{sec:model}.\\ Consider what is arguably the simplest implementation of the above idea: The DM $\chi$ is a Majorana particle that annihilates into a SM right-handed quark $q$ and a heavy vector-like antiquark $\bar \Psi$ (and into the CP conjugate states). Moreover, $\Psi$ decays into the Higgs and a quark, conserving $B-L$, but the Yukawa couplings are taken very small, so that most of the $\Psi$'s decay after the sphalerons freeze out at $T=T_{\rm sfo}$. The lepton asymmetry is also frozen at $T_{\rm sfo}$ because the sphalerons are the only processes violating $L$ (here $L=L_{\rm SM}$), with the freeze-in $L$-asymmetry of the order of the $\Psi$-asymmetry at $T=T_{\rm sfo}$. As we explained before, the baryon asymmetry today has to be equal to this frozen lepton asymmetry, which is in general not null. However this model badly fails to generate the observed BAU. The issue is related to the washouts and can be understood through a simple analysis of the terms in the BE. The most dangerous one reads [see Eq.~\eqref{eq:be4}]:
\begin{equation} \label{eq:dangwo}
3\,s z H \frac{\dif Y_{B_{SM}-L}}{\dif z} = 2 \frac{Y_\Psi - Y_{\bar \Psi}}{Y_{\Psi}^{eq}} \ghor{\bar U_R \Psi}{\bar \Psi U_R} + \dots ,
\end{equation}   
where $s$ is the entropy density, $z\equiv m_\chi/T$, $H$ is the Hubble rate, and $\gamma$ is a reaction density.
The crucial point is that the effect of the washout term on the r.h.s. depends on how the asymmetry $Y_\Psi - Y_{\bar \Psi}$ is related to $Y_{B_{SM}-L}$. In our case, $Y_\Psi - Y_{\bar \Psi} = - Y_{B_{SM}-L}$, because there is a conservation law that involves only the SM fields and $\Psi$. Then we can rewrite that washout term as
\begin{equation}\label{eq:noBsuppression}
3\,s z H \frac{\dif Y_{B_{SM}-L}}{\dif z} =  -2\, Y_{B_{SM}-L} \frac{\ghor{\bar U_R \Psi}{\bar \Psi U_R}}{Y_{\Psi}^{eq}} + \dots 
\end{equation}  
Since $\frac{\ghor{\bar U_R \Psi}{\bar \Psi U_R}}{Y_{\Psi}^{eq}}$ is not Boltzmann suppressed, this washout badly erases the asymmetry, independently of how heavy $\Psi$ is.

We have arrived at this conclusion exemplifying with a model where $L_{SM}$ is conserved  pertubatively, but similar considerations hold trading  $L_{SM}$ for  $B_{SM}$ . The bottom line is that in models with a conserved $B-L$, there must be other fields besides $\Psi$ and the SM ones. Nevertheless, as we will demonstrate in Sec.~\ref{sec:model}, the additional particles need not be stable and constitute a sequestered sector in the universe today, but instead can be heavy and decay entirely into SM particles. 

\item[(b)] {\it B-L violating case}: We mentioned above that in the $B-L$ conserving models proposed in Ref.~\cite{Cui:2011ab} the authors argue for the necessity of a light sterile sector and a $\mathbb{Z}_4$ symmetry. Those arguments do not apply to the $B-L$ violating case. Even so, {\em all} the models presented in Ref.~\cite{Cui:2011ab} or included in the effective approach of Ref.~\cite{Bernal:2012gv} possess such a sterile sector and discrete asymmetry. For example there are models~\cite{Cui:2011ab,Bernal:2012gv} in which $B$ is violated at the perturbative level, but still $\Psi$ decays into two quarks plus a SM light singlet, and again a $\mathbb{Z}_4$ symmetry is imposed to avoid $\Psi$ disintegration entirely into SM baryons (together with other processes that can washout the asymmetry). In addition, proton stability is guaranteed by the $\mathbb{Z}_4$. It is interesting to show with a concrete example that also in models with violation of $B-L$ it is possible to generate the BAU from DM annihilations, without light sterile particles and invoking just a $\mathbb{Z}_2$ symmetry. We do this in Sec.\ref{sec:modelC}, presenting a leptogenesis model which can also yield some relation to neutrino masses.    
\end{itemize}


\section{A model with $B-L$ conservation} \label{sec:model}

\begin{figure}[!t]
\begin{center}
\includegraphics[width=1.0\textwidth]{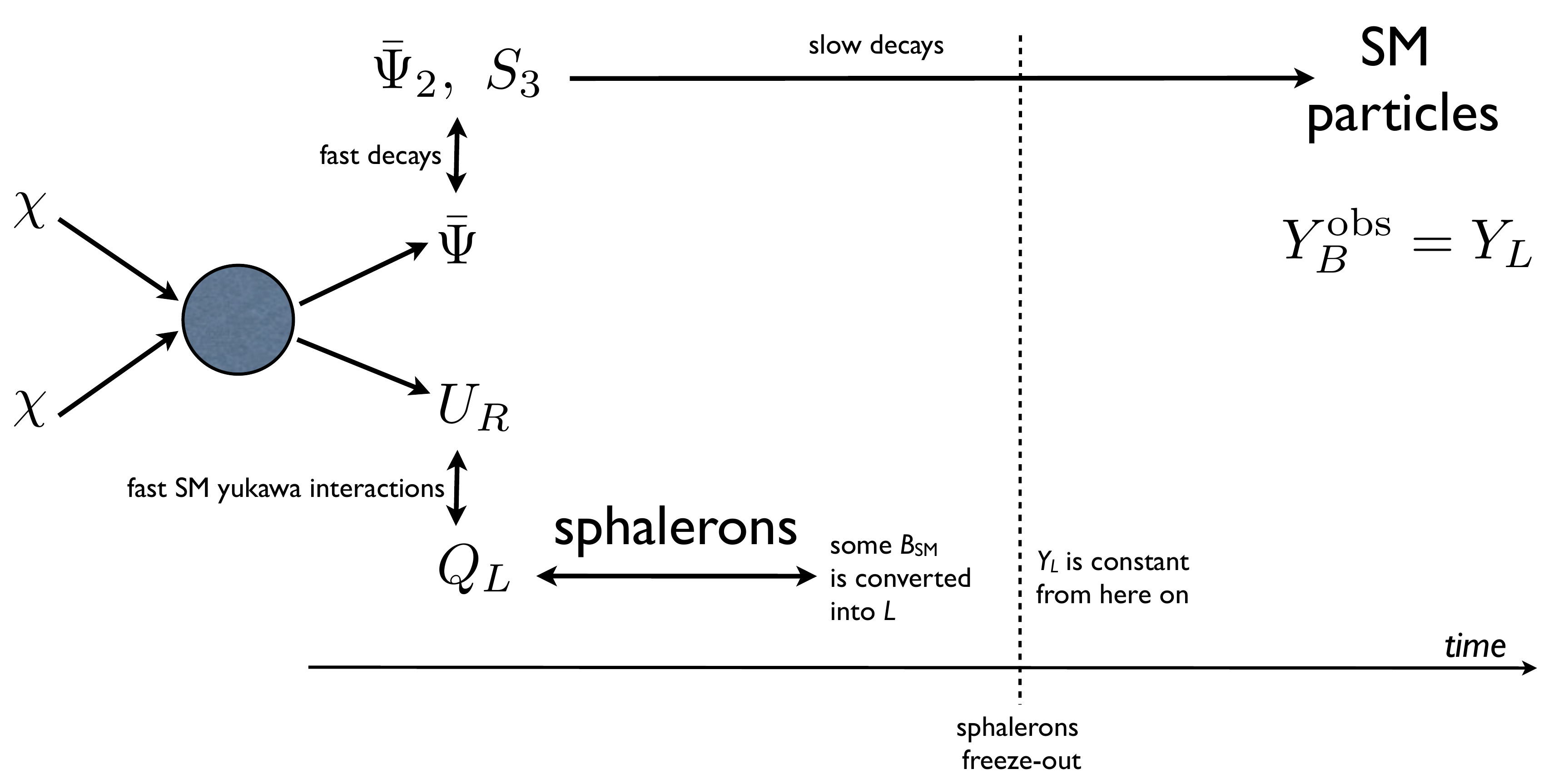}
\end{center}
\vspace{-0.8cm}
\caption{\sl \textbf{\textit{Cartoon of the mechanism of the model with $B-L$ conservation.}} 
}
\label{Fig:sketch}
\end{figure}
We will show in this section that one can generate a baryon asymmetry via dark matter annihilation without the need for a light stable dark sector particle and without imposing a $\mathbb{Z}_4$ symmetry.\footnote{The only dark sector particle is the WIMP, which is heavy.}

The idea is schematically shown in Fig.~\ref{Fig:sketch}. Dark matter annihilates into a SM quark, $U_R$, and an exotic heavy antiquark, $\bar \Psi$, both singlets under the SM weak gauge group $SU(2)$, and into their CP conjugate states, $\bar U_R$ and $\Psi$. The annihilation is $CP$ violating and generates an asymmetry in $B_{\rm SM}$ and $B_\Psi$. Fast SM Yukawa interactions convert $U_R$ into left-handed quarks, $Q_L$, and sphalerons act on $Q_L$, converting part of $B_{\rm SM}$ into lepton number, $L=L_{\rm SM}$. Meanwhile $\Psi$ decays quickly into $\Psi_2$ and $S_3$, with $\Psi_2$ a fermion with the same quantum numbers as $\Psi$, and $S_3$ a scalar singlet. When the sphalerons freeze-out the $L$ asymmetry is frozen, and some time after that $\Psi_2$ decays entirely into SM particles. The net result is that we are left with a baryon asymmetry $Y_{B}^{\rm obs}=Y_{L}$ (with $Y_{B}^{\rm obs}$ the observed BAU). Since the heavy exotic particles have all decayed into SM particles, there is no trace of them in the present universe.

In other words, what we are doing here is essentially the following: instead of demanding a stable light dark sector to contain a baryon asymmetry opposite to that of the visible universe, as done in Ref.~\cite{Cui:2011ab}, the ``negative'' asymmetry is momentarily stored in some heavy fields, which decay after sphalerons freeze out, allowing for a sizable $L$-asymmetry to get frozen together with the sphalerons. At the end, the BAU is equal to this lepton asymmetry.

A few comments are in order.
\begin{itemize}
\item The DM annihilations violate $B_{SM}$ and $B_{\Psi}$, but they conserve $B$ and $L$. Moreover, the electroweak sphalerons conserve $B_{SM}-L$, $B_{\Psi}$ and $B_{\Psi_2}$, because $\Psi$ and $\Psi_2$ are vector-like. Then it is clear that $B-L \equiv (B-L)_{\rm SM} + (B-L)_\Psi + (B-L)_{\Psi_2}$ is conserved in this model. For the mechanism we are proposing, it is also important to stress that $L$ is only violated by the sphalerons, and therefore remains constant after they freeze out. 
\item Why do we need $\Psi_2$ and $S_3$? It has to do with the issue outlined in the previous section. If $\Psi$ decayed into $HQ_L$ through the Yukawa $\lambda_\Psi H \bar Q_L P_R \Psi$, we would run into the problem encoded in Eq.~\eqref{eq:noBsuppression}. A solution to this problem is to open a new, dominant decay channel, $\Psi \to \Psi_2\,S_3$. This brings a significant change: the condition $Y_\Psi - Y_{\bar\Psi} + Y_{B_{\rm SM}-L} = 0$ is replaced by the following
\beqn
Y_{\Delta \Psi} + Y_{\Delta \Psi_2} + Y_{B_{\rm SM}-L} &=& 0 \; ,\\
\mu_\Psi &=& \mu_{\Psi_2} \; ,
\eeqn
with $\mu$ the chemical potential, and $Y_{\Delta X} \equiv Y_X - Y_{\bar X}$. As a consequence, $Y_{\Delta \Psi}$ on the right hand side of Eq.~\eqref{eq:dangwo} is replaced by 
\beqn \label{eq:Boltzsup}
Y_{\Psi - \bar \Psi} &=&- \frac{m_{\Psi}^2\,K_2(m_\Psi / T)}{m_{\Psi}^2\,K_2(m_\Psi / T)+m_{\Psi_2}^2\,K_2(m_{\Psi_2}/ T)} \; Y_{B_\text{SM}-L} \nonumber \\
  &\simeq & - \; \frac{(m_{\Psi}/m_{\Psi_2})^{3/2} \; \myexp{-(m_{\Psi}-m_{\Psi_2})/T}}{1 +(m_{\Psi}/m_{\Psi_2})^{3/2} \; \myexp{-(m_{\Psi}-m_{\Psi_2})/T} } \; Y_{B_\text{SM}-L} \; ,
\eeqn
where $K_n$ are the modified Bessel functions of the $n^\text{th}$-type, and the second line is valid in the limit $(m_\Psi - m_{\Psi_2}) \gg T$.
We give more details of the derivation of this result in Appendix~\ref{app:BEs}. The key point is that now the washout term in Eq.~\eqref{eq:dangwo} is proportional to $\myexp{-(m_\Psi-m_{\Psi_2})/T}$, and hence it is Boltzmann suppressed as long as $m_{\Psi_2} \ll m_{\Psi}$. This results in the fact that washout processes will decouple before DM freeze-out, satisfying one of the crucial conditions to successfully achieve WIMPy baryogenesis.
\item The couplings of $\Psi_2$ to SM quarks must be small enough to have $\Psi_2$ decay after sphalerons freeze out and to suppress washout processes like $\lrproname{\bar \Psi_2 U_R}{\Psi_2 \bar U_R}$. Another key point is that these couplings can be taken tiny without reducing the CP asymmetry, which is proportional to the couplings of $\Psi$, not $\Psi_2$.  
\end{itemize}

We are now in the position to write a Lagrangian that exemplifies how the mechanism works.

\subsection{Lagrangian and parameters}
\begin{table}[!t]
\centering
\begin{tabular}{ c c c c c c}
\toprule
{} & $SU(3)$ & $SU(2)_L$ & $Q_{U(1)_y}$ & $Q_{U(1)_B}$ &  $\mathbb{Z}_2$ \\
\midrule
$\chi$ & 1 & 1 & 0 & 0 & $-1$ \\
$\Psi$ & 3 & 1 & $+2/3$ & $+1/3$ & $+1$ \\
$\Psi_2$ & 3 & 1 & $+2/3$ & $+1/3$ & $+1$ \\
$Q_L$ & $3$ & 2 & $+1/6$ & $+1/3$ & $+1$ \\
$U_R$ & $3$ & 1 & $+2/3$ & $+1/3$ & $+1$ \\
\midrule
$S_{1,2,3}$ & $1$ & 1 & $0$ & $0$ & $+1$ \\
$H$ & $1$ & 2 & $+1/2$ & $0$ & $+1$ \\
\bottomrule
\end{tabular}
\caption{Particle content of the model with $B-L$ conservation.} \label{Tab:contentB}
\end{table}

We list the particle content of the model, with the corresponding SM quantum numbers, in Table~\ref{Tab:contentB}.
The particles in the first block are fermions: $\chi$, the WIMP, is Majorana\footnote{Note that in the models of Refs.~\cite{Cui:2011ab, Bernal:2012gv} the DM had to be Dirac because it had a complex charge under the $\mathbb{Z}_4$. Instead in our model it {\em can} be Majorana.}; $\Psi$ and $\Psi_2$ are exotic heavy vector-like quarks; $U_R$ is one of the right-handed up-type SM quarks, $Q_L$ is the corresponding left-handed quark (for simplicity $\Psi$ couples only to one flavor of quarks in this model). In the second block $S_1$,  $S_2$ are pseudoscalars, while $S_3$ can be either a scalar or a pseudoscalar; $H = \begin{pmatrix} H^+ \\ H^0 \end{pmatrix}$ is the Higgs doublet and we use the notation $\tilde H \equiv i \tau_2 H^* = \begin{pmatrix} H^{0\,*} \\ -H^- \end{pmatrix}$, with $\tau_2$ the second Pauli matrix. The $\mathbb{Z}_2$ is imposed to make the DM stable.
The Lagrangian is
\beqn
L &=& L_{SM} + L_{kin} + V(S_i,H) \nn \\
&+&\frac{1}{2} m_\chi \bar\chi^c \chi + m_\Psi \bar\Psi \Psi  + m_{\Psi_2} \bar\Psi_2 \Psi_2+ \frac{1}{2} m_{S_\alpha}^2 S_\alpha^2  + \frac{1}{2} m_{S_3}^2 S_3^2 \nn \\
&+& i \lambda_{\chi_\alpha} S_\alpha \bar\chi^c \gamma_5 \chi + i \lambda_{B_\alpha} S_\alpha  \bar U P_L \Psi \nn \\
&+& \lambda_3 S_3 \bar\Psi \Psi_2 + \lambda_\Psi  \bar Q \tilde H P_R \Psi + \lambda_{\Psi_2} \bar Q \tilde H P_R \Psi_2 + {\rm h.c.} \label{eq:Lagcouplings},
\eeqn
with $P_{R,L} =  (1 \pm \gamma_5)/2$.
Some comments are in due order. The index $\alpha = 1,2$ runs over the first two families of pseudscalars, while $i=1,2,3$ runs over all of them. We assume for simplicity that in the scalar potential $V(S_i, H)$ there are no sources of $CP$ violation and the couplings between the new scalars $S_i$ and the Higgs are very small. The fields $\Psi, \Psi_2, S_{1,2,3}$ are mass eigenstates, so we are assuming that we have already diagonalized the corresponding mass matrices. 
 Note that below the EW scale, $v$, one would have to diagonalize the mass matrix for the scalars including the Higgs and the one for the fermions including the SM quarks, but given the hierarchy $m_\Psi, m_{\Psi_2},m_{S_{1,2,3}} \gg v$ and the tiny couplings\footnote{We will explain in the next subsection why these couplings have to be tiny.} $\lambda_\Psi,\lambda_{\Psi_2}$, the mass eigenstates  $\Psi, \Psi_2, S_{1,2,3}$ would remain such to a good approximation and have small mixings with SM particles. 
$S_1$ and $S_2$ mediate the DM annihilation. We choose them to be pseudoscalars so that the annihilation is not velocity suppressed. We need at least two of them in order to have a physical, rephasing invariant CP phase, a condition necessary to generate a CP asymmetry. $S_3$, instead, is needed for the decay $\Psi \to \Psi_2 S_3$, but to keep it simple we do not want it to mediate the DM annihilation, so we consider the couplings $i \lambda_{\chi_3} S_3 \bar\chi \gamma_5 \chi$ and $i \lambda_{B_3} S_3  \bar U P_L \Psi$ to be negligible. 
 
According to the symmetries, there are some more terms that we could add to the Lagrangian. We will assume that such extra terms have small couplings and can be neglected. This way we can focus just on the couplings that we have written and illustrate the mechanism in the simplest possible way.


\subsection{CP asymmetry, washouts and decays}
\begin{figure}[!t]
\begin{center}
\includegraphics[width=1.0\textwidth]{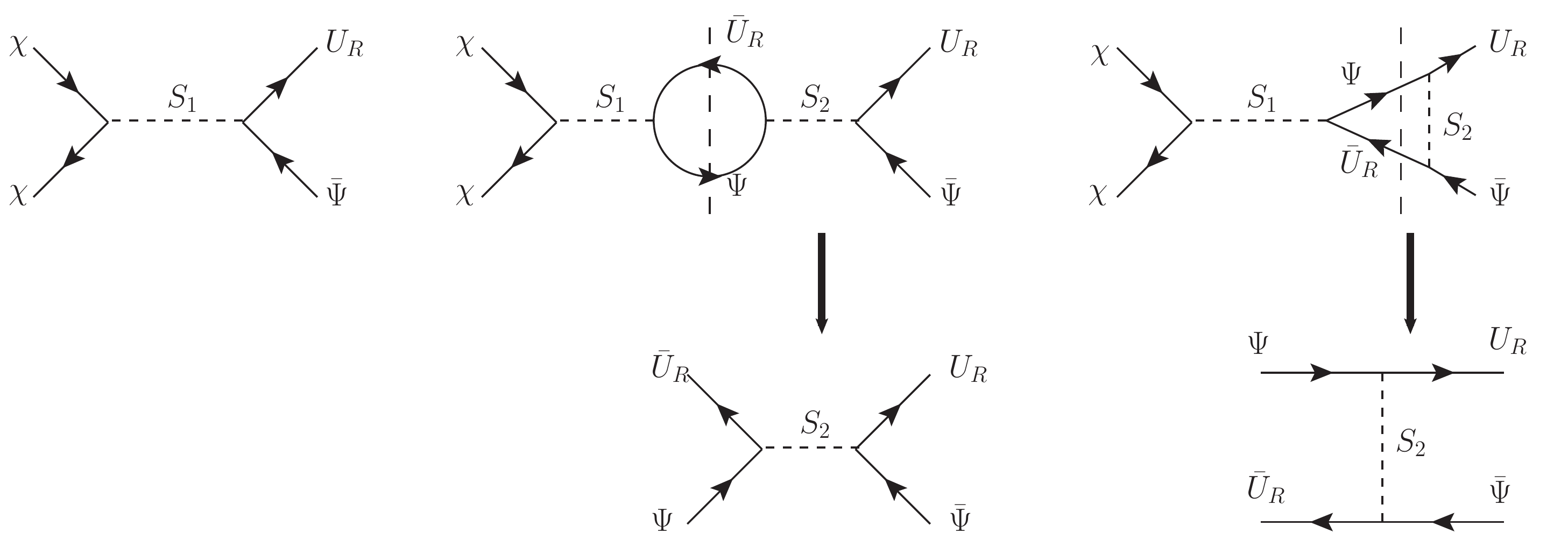}
\end{center}
\vspace{-0.8cm}
\caption{\sl \textbf{\textit{CP asymmetry and washout processes.}} We show on the top a combination of tree-level and one-loop diagrams that lead to a CP asymmetry. There are more diagrams to include in the full calculation, the result of which can be found in Appendix~\ref{app:crossNico}. The vertical cuts through the loops indicate that those particles can go on shell, which is necessary to get a kinematical phase. In turn, this implies the existence of the tree level diagrams on the bottom, which contribute to the washout. 
}
\label{Fig:CPandwashout}
\end{figure}

The CP asymmetry arises via the interference between tree-level and one-loop diagrams, as we illustrate in Fig.~\ref{Fig:CPandwashout}.

The relevant washout processes are the same as in 
Ref.~\cite{Cui:2011ab}. They all involve the heavy $\Psi$. In principle one has to worry about $\Psi_2$ as well, but in our model it contributes to the washouts only through $3 \leftrightarrow 3$ processes of the type $S_3 \Psi_2 \bar U_R \leftrightarrow S_3 \bar \Psi_2 U_R$. These are phase-space suppressed and we have verified that they can safely be neglected.

\begin{figure}[!t]
\begin{center}
\includegraphics[width=0.4\textwidth]{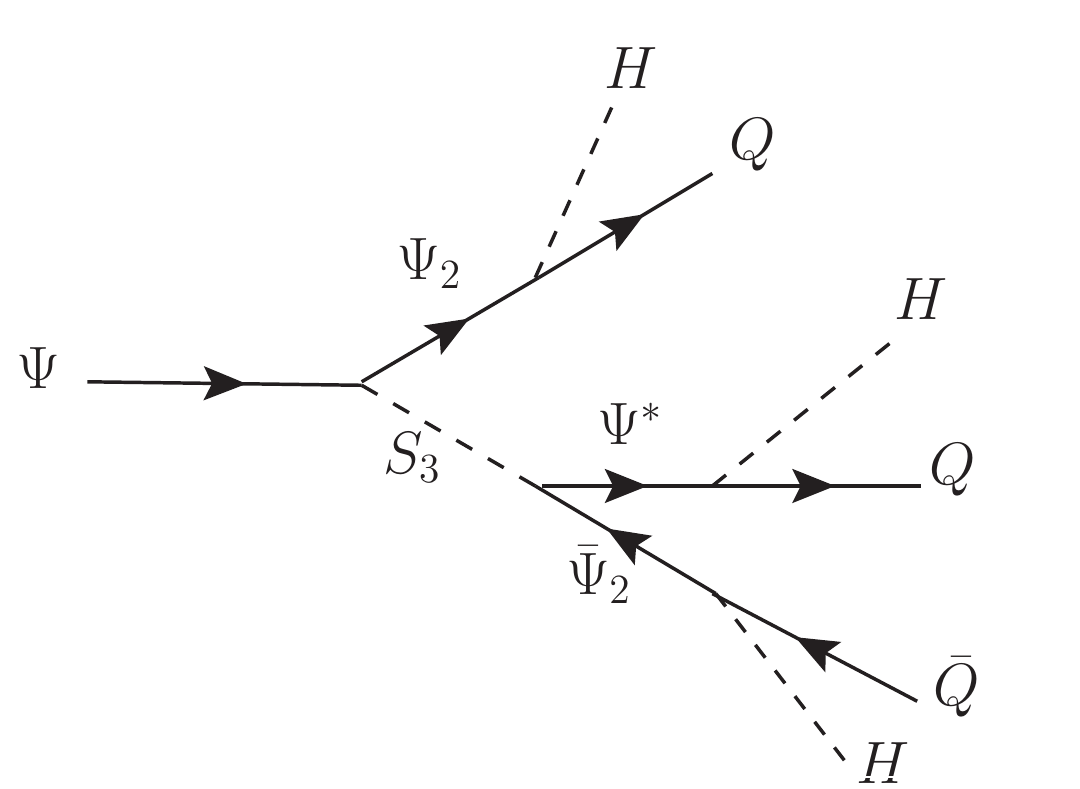}
\end{center}
\vspace{-0.8cm}
\caption{\sl \textbf{\textit{Decay chain for the exotic heavy quark.}} At the end of the decay chain we are left with only SM particles. $\Psi^*$ in the internal line indicates that the particle is off shell.
}
\label{Fig:Decaychain}
\end{figure}

As we have mentioned, $\Psi$ stays in thermal equilibrium during DM annihilation thanks to the fast decays into $\Psi_2 S_3$. In turn, $\Psi_2$ decays into $H Q_L$ after sphaleron decoupling. We have the following constraints on the couplings that appear in Eq.~\eqref{eq:Lagcouplings}: 
\begin{itemize}
\item[-] $\lambda_3$ has to be much bigger than $\lambda_\Psi$ for the decay $\Psi \to \Psi_2 S_3$ to be dominant over $\Psi \to H Q$.
\item[-] $\lambda_{\Psi_2} \lesssim 10^{-7}$ to guarantee that $\Psi_2$ decays after sphalerons freeze out.
\end{itemize} In Fig.~\ref{Fig:Decaychain} we show the whole decay chain from $\Psi$ to SM particles. The kinematics of the annihilation and the decays just described require the following mass hierarchy:
\be
2\,m_\chi > m_\Psi > m_{S_3} + m_{\Psi_2},\qquad\qquad m_{S_3} > m_{\Psi_2} > 0.8 \ {\rm TeV}.
\ee
The last inequality is dictated by LHC bounds on vector-like quarks, as we will explain in more detail in a following subsection dedicated to experimental constraints. 
Note also that the washouts involving an external $S_3$, $S_3 \leftrightarrow \bar\Psi_2 HQ$ and $S_3 \Psi_2 \leftrightarrow HQ$, are suppressed with respect to $\Psi \leftrightarrow HQ$, if $m_{S_3}$ is taken not much smaller than $m_\Psi$.

We report all the relevant cross sections and decay rates in Appendix~\ref{app:crossNico}.

\subsection{Numerical results}
The Boltzmann Equations (BEs) are derived in detail in Appendix~\ref{app:BEs}. In this section we report the numerical results obtained for two benchmarks. In the first one we set the masses of the mediators $S_1$ and $S_2$ to high values, 20 and 22 TeV respectively. However with heavy mediators we need bigger couplings, approaching the $4\pi$ bound from perturbativity, to obtain cross sections and CP asymmetries that lead to the right DM abundance and BAU. In the second benchmark we use lower masses for $S_1$ and $S_2$, which allow for lower couplings, but require more care when dealing with the propagators because there are subtle issues related to broad resonances. We discuss this in more detail in Appendix~\ref{app:crossNico}.

We do not perform a scan of the parameter space of the model. For such a scan to be manageable, we would have to make further assumptions and impose extra conditions to reduce the number of parameters. However, our aim here is simply to show that the mechanism we are proposing can indeed generate the observed BAU, while giving the correct WIMP abundance. The value of the relevant parameters for the benchmarks are shown in Table~\ref{Tab:bench}. $m_{S_3}$ and $\lambda_3$ do not appear in the table because the outcome of the BEs is almost insensitive to their exact values,
 as long as they satisfy the conditions $\lambda_3 \gg \lambda_\Psi$ and $m_{\Psi_2} < m_{S_3} < m_\Psi - m_{\Psi_2}$, that we mentioned above.

\begin{table}[!t]
\centering
\begin{tabular}{ l|ccccc|cccc l }
\toprule
Parameter & $m_{\chi}$ & $m_{\Psi}$& $m_{\Psi_2}$ & $m_{S_1}$& $m_{S_2}$  & $\lambda_{\chi_1}$ & $\lambda_{\chi_2}$ & $\lambda_{B _1}$ & $\lambda_{B_2}$\\
\midrule
Benchmark A & 4 & 6.7 & 0.8 & 20 & 22  & 2.9 & 3.2 & 3 & 3.3\\
\midrule
Benchmark B & 2 & 3.1 & 0.8 & 5 & 5.5  & 0.50 & 0.55 & 0.70 & 0.77\\
\bottomrule
\end{tabular}
\caption{Benchmark points for the model with $B-L$ conservation. The masses are in TeV. Additionally, we set, $\lambda_\Psi=5\times 10^{-4}$ and $\lambda_{\Psi_2}=5\times 10^{-8}$ for both benchmarks. The CP-violating phase $\alpha$, defined by $\lambda_{B 1}\lambda_{B 2}^* \equiv |\lambda_{B 1}| |\lambda_{B 2}| e^{i\alpha}$, is fixed to $2 \pi / 3$ to maximize the asymmetry.} \label{Tab:bench}
\end{table}

\begin{figure}[!h]
\begin{center}
\includegraphics[width=0.8\textwidth]{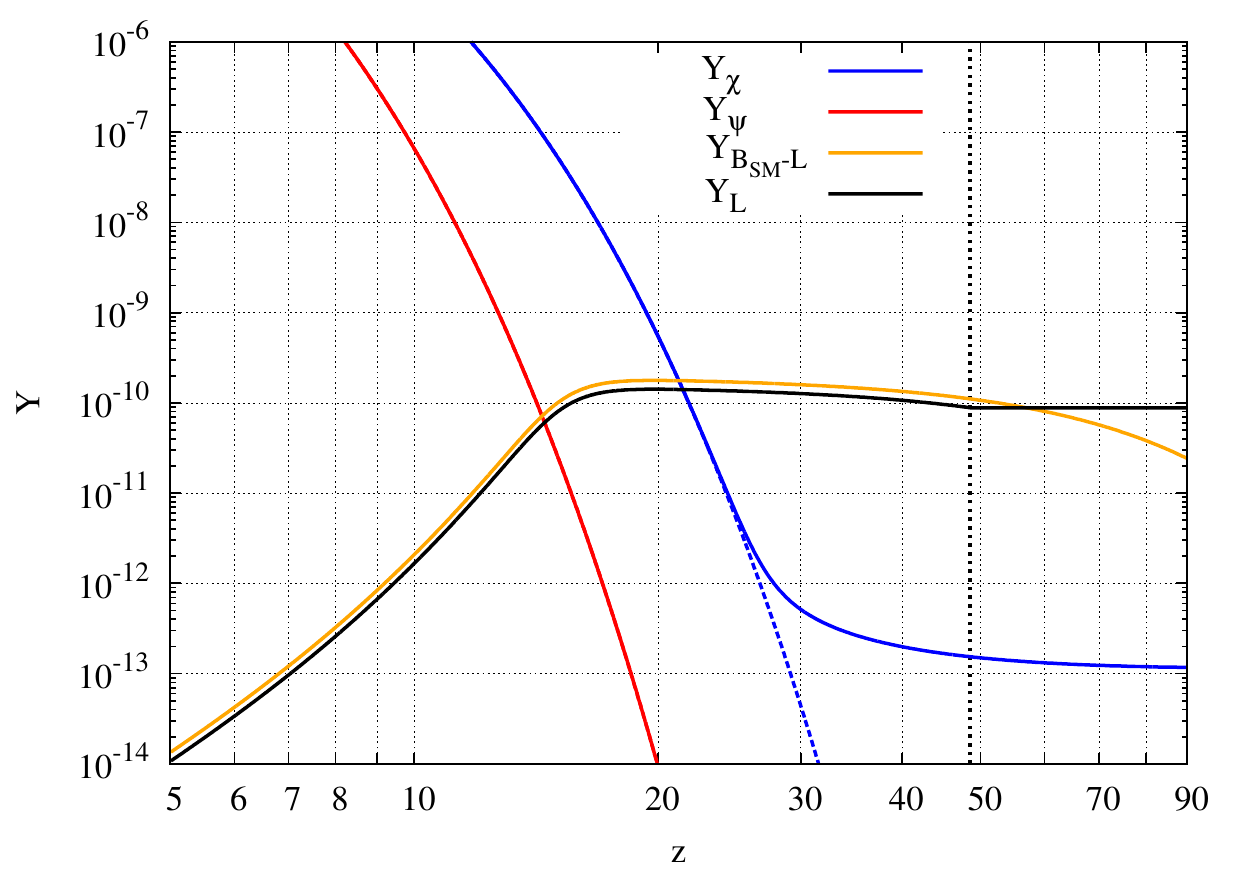}
\end{center}
\vspace{-0.8cm}
\caption{\sl \textbf{\textit{Benchmark A for the $B-L$ conserving model.} } We show $Y = n/s$ as a function of $z=m_\chi/T$ for the DM, the exotic field $\Psi$, and the asymmetries $Y_{B_{\rm SM}-L}$ and $Y_L$. The blue dashed line traces the DM equilibrium distribution. The vertical black dotted line denotes the approximate point at which sphalerons freeze out. The BEs to get these curves can be found in Appendix~\ref{app:BEs}. The parameters used for this benchmark point are listed in Table~\ref{Tab:bench}. 
}
\label{Fig:Nico1A}
\end{figure}

\begin{figure}[!h]
\begin{center}
\includegraphics[width=0.8\textwidth]{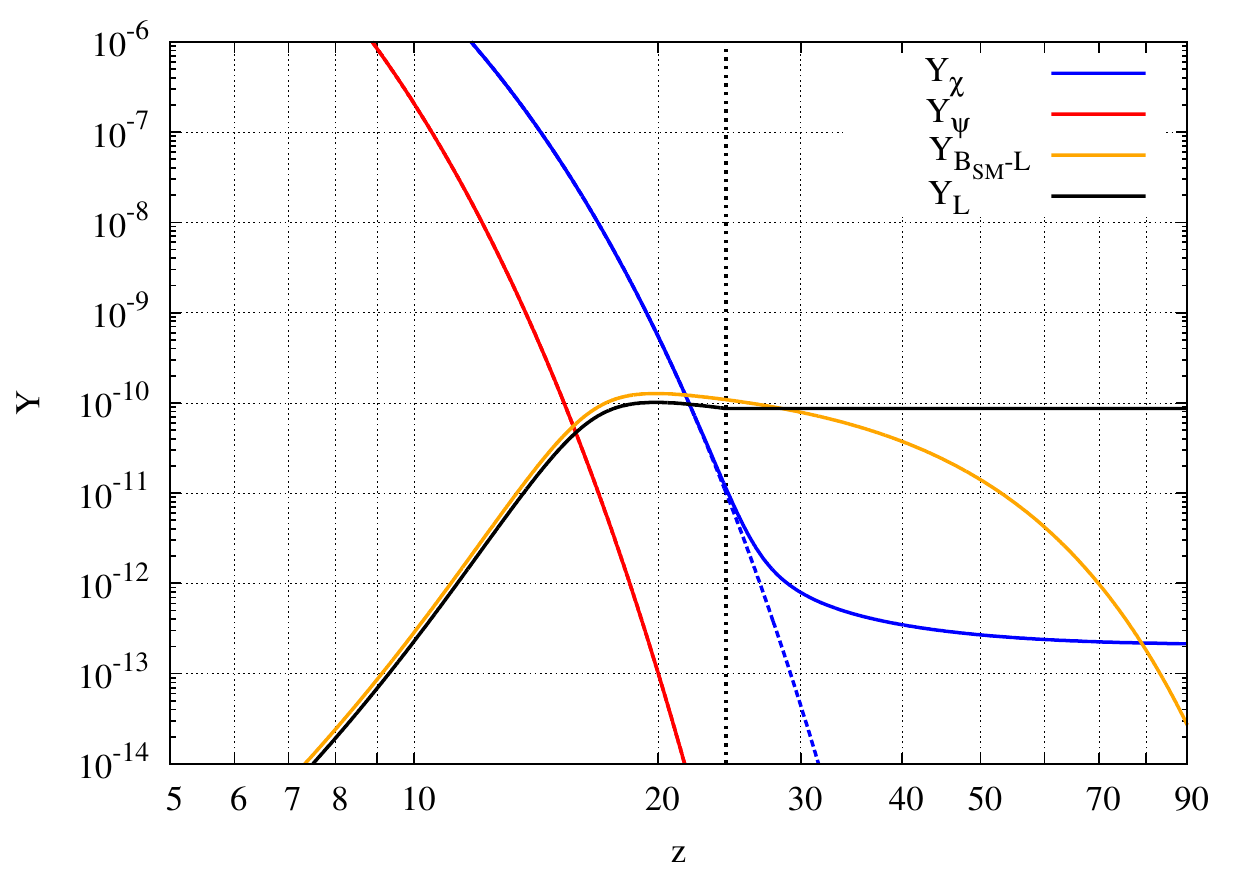}
\end{center}
\vspace{-0.8cm}
\caption{\sl \textbf{\textit{Benchmark B for the $B-L$ conserving model.}} See caption of Fig.~\ref{Fig:Nico1A}. 
}
\label{Fig:Nico1B}
\end{figure}

We show the evolution of the densities in Figs.~\ref{Fig:Nico1A}, \ref{Fig:Nico1B}. In the plots one can see the following features. $\Psi$ (red curve) is in thermal equilibrium thanks to the fast processes $\Psi \leftrightarrow \Psi_2 S_3$. The DM (blue curve) freeze-out occurs at values of $z\equiv m_\chi / T$ between 20 and 30, as is typical of the WIMP scenario. The black curve for the lepton asymmetry, $Y_L$, follows closely the orange one for $Y_{B_{\rm SM}-L}$. The two quantities are related by the condition $Y_{L} = - \frac{63}{79} Y_{B_{\rm SM}-L}$  [see Eq.~\eqref{eq:Yrel}] for $z<z_{\rm sfo}$, with $z_{\rm sfo}\equiv m_\chi / T_{\rm sfo}$ corresponding to the freeze-out of sphalerons that we take to be at $T_{\rm sfo} \sim T_c/1.7$~\cite{Pilaftsis:2008qt}, where $T_c = 140$ GeV is the critical temperature for the electroweak phase transition~\cite{Burnier:2005hp,D'Onofrio:2012ni}. For $z > z_{\rm sfo}$ the lepton asymmetry $Y_L$ is constant and when all the exotic fields have decayed we get $Y_{B_{\rm SM}-L}=0$ and a baryon asymmetry $Y_{B_{\rm SM}} = Y_L$ that matches the observed value.

\subsection{Experimental constraints}
The mechanism we are proposing involves new colored particles, $\Psi$ and $\Psi_2$, which are subject to constraints from the LHC. Searches for vector-like quarks~\cite{ATLAS-CONF-2013-018} put a lower bound on the mass of these objects between 650 and 800 GeV. To be conservative, we will take the higher value and require $m_\Psi> m_{\Psi_2} > 800$ GeV. 
We showed in Eq.~\eqref{eq:Boltzsup} that a crucial ingredient of our model is the Boltzmann suppression, proportional to $e^{-(m_\Psi-m_{\Psi_2})/T}$, of the washouts. The lower bound of 800 GeV on $m_{\Psi_2}$ then pushes the mass of $\Psi$ up to a few TeV for the Boltzmann suppression to be effective. In turn, the DM mass has to satisfy the kinematical bound $2m_\chi > m_\Psi$.
 Note also that for the models with DM annihilating into a quark and an exotic heavy antiquark studied in Refs.~\cite{Cui:2011ab, Bernal:2012gv}, LHC bounds came from searches with missing energy. That was a consequence of the decay of $\Psi$ into quarks plus a light sterile dark sector particle (the missing energy). Those bounds are not relevant to the model studied in this paper, given that the decays of $\Psi$ and $\Psi_2$ are different here.

At the LHC one could also probe the DM directly, rather than the vector-like quarks, via monojet searches~\cite{Chatrchyan:2012me,ATLAS:2012ky}. However this does not seem promising for our models for two reasons. First, monojet searches are typically more competitive than direct detection experiments when the DM is lighter than a few tens of GeV. Our DM candidate is much heavier. Second, the process would be loop-suppressed because a pair of our DM particles couples to a SM quark and a vector-like quark instead of a pair of SM quarks.

We find that there are no bounds from DM direct detection experiments. Given the interactions in this model, the lowest order contribution to a direct detection cross section would naively be one-loop~\cite{Cui:2011ab}. It turns out, as pointed out in Ref.~\cite{Bernal:2012gv}, that due to cancellations this contribution is even further suppressed, which unfortunately leaves almost no prospect of detection in the near future.

The best constraints on the CP-violating phases, needed for the generation of the BAU, would come from Electric Dipole Moment (EDM) measurements. In Ref.~\cite{Cui:2011ab} the authors pointed out that, when the new heavy fields couple only to one chirality of light SM fermions, the lowest order contribution to EDMs occurs at three loops, thus it is very suppressed. In our model, $\Psi$ couples to right-handed quarks via the pseudoscalars $S_\alpha$, but also to left-handed quarks via the couplings in the last line of Eq.~\eqref{eq:Lagcouplings}. It is then possible that there are two-loop diagrams that contribute to EDMs, which contain the couplings $\lambda_{B_\alpha}, \lambda_\Psi, \lambda_{\Psi_2}$ and are proportional to the relative phase of $\lambda_{B_1}$ and $\lambda_{B_2}$ that is responsible for the CP asymmetry. However the couplings $\lambda_\Psi$ and $\lambda_{\Psi_2}$ are tiny and provide a strong suppression. Investigating the structure of these diagrams is beyond the scope of the current work, but in light of these considerations we can state that EDM bounds do not pose a serious challenge to this model. 


\section{A model with $B-L$ violation} \label{sec:modelC}

\begin{figure}[!t]
\begin{center}
\includegraphics[width=1.0\textwidth]{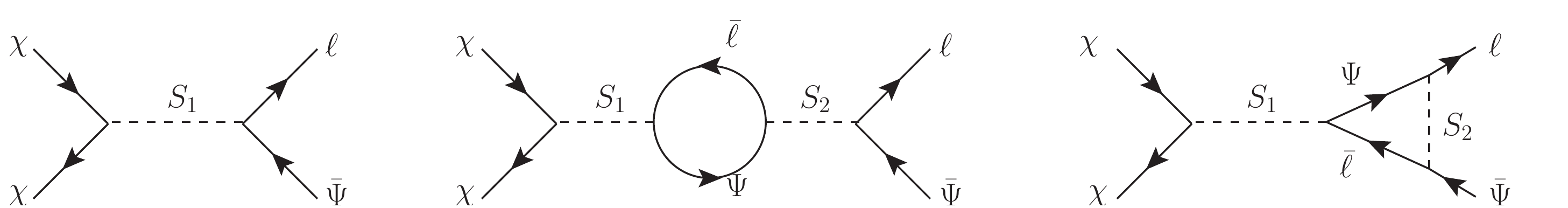}
\end{center}
\vspace{-0.8cm}
\caption{\sl \textbf{\textit{Origin of the CP asymmetry in the model with $B-L$ violation.}} We show here, as an example, only some combinations of diagrams that generate an asymmetry. }
\label{Fig:lepasym}
\end{figure}

\begin{figure}[!t]
\begin{center}
\includegraphics[width=0.4\textwidth]{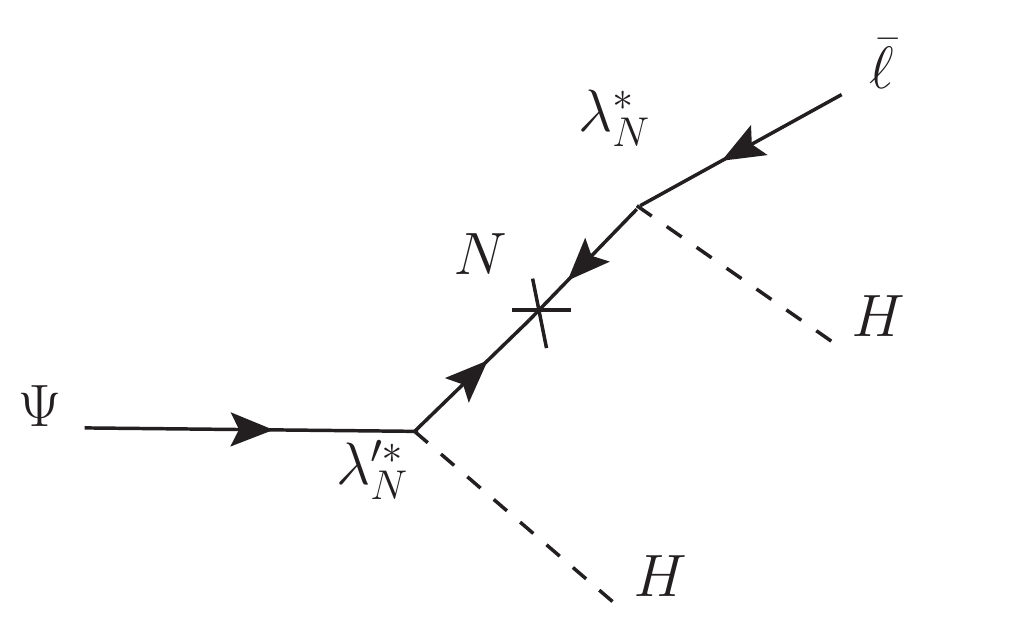}
\end{center}
\vspace{-0.8cm}
\caption{\sl \textbf{\textit{Decay of the heavy vector-like lepton.}} This decay is $L$-violating. The final products only contain SM particles.}
\label{Fig:PsiLdecay}
\end{figure}

In this section we present a leptogenesis model with $B-L$ violation. Dark matter annihilates into a SM lepton doublet, $\ell$, and an exotic heavy vector-like doublet antilepton, $\bar \Psi$. An asymmetry again arises as shown in Fig.~\ref{Fig:lepasym}. Then we introduce a heavy singlet neutrino, $N$, that opens up an $L$-violating decay channel for $\Psi$, via a Yukawa coupling of the type $\lambda'_{N} \bar\Psi \tilde H N$, as shown in the diagram of Fig.~\ref{Fig:PsiLdecay}. This is in contrast with the leptogenesis model proposed in Ref.~\cite{Cui:2011ab}, where a similar Yukawa coupling,  $\lambda_{n} \bar\Psi \tilde H n$, was present but $n$ was massless and stable, and a $\mathbb{Z}_4$ symmetry was imposed, under which both $\Psi$ and $n$ were charged, to forbid the term $\lambda'_{e} \bar\Psi H e_R$. Instead in our model we impose a $\mathbb{Z}_2$, under which only the DM is charged so that it is stable, and we assume that the Yukawa $\lambda'_{e}$ is small enough, so that the $L$-conserving decay channel $\Psi \to H\,e_R$ is subdominant compared to the $L$-violating one $\Psi \to H\,N$.

In the simple example just outlined we introduced only one singlet neutrino, but it is immediate to add more and explain neutrino masses, as we discuss in Sec.~\ref{neutrinomasses}.

\subsection{Lagrangian and parameters}

\begin{table}[!t]
\centering
\begin{tabular}{ c c c c c}
\toprule
{} & $SU(3)$ & $SU(2)_L$ & $Q_{U(1)_y}$ &  $\mathbb{Z}_2$ \\
\midrule
$\chi$ & 1 & 1 & 0  & $-1$ \\
$\Psi$ & 1 & 2 & $-1/2$ & $+1$ \\
$N_i$ & 1 & 1 & 0 & $+1$ \\
$\ell$ & $1$ & 2 & $-1/2$ &  $+1$ \\
$e_R$ & $1$ & 1 & $-1$ & $+1$ \\
\midrule
$S_{1,2}$ & $1$ & 1 & $0$ & $+1$ \\
$H$ & $1$ & 2 & $+1/2$  & $+1$ \\
\bottomrule
\end{tabular}
\caption{Particle content of the model with $B-L$ violation.} \label{Tab:contentC}
\end{table}

We consider the particle content shown in Table~\ref{Tab:contentC}. The fields $\chi, S_{1,2}$ are replicas of those employed in the model of Section~\ref{sec:model};  $\Psi$ is a heavy vector-like lepton doublet; $N_i$ are singlet heavy neutrinos, with $i=1,\dots,n$; $\ell$ and $e_R$ are the SM $SU(2)$ lepton doublet and singlet respectively; $H$ is the SM Higgs doublet. The Lagrangian is
\beqn
L & \supset & \frac{1}{2} m_\chi \bar\chi^c \chi + m_\Psi \bar\Psi \Psi  + \frac{1}{2} m_{S_\alpha}^2 S_\alpha^2 +\frac{1}{2} \bar N^c_i m_N^{ij} N_j \nn \\
&+& i \lambda_{\chi_\alpha} S_\alpha \bar\chi^c \gamma_5 \chi + i \lambda_{L_\alpha} S_\alpha  \bar \Psi  \ell  \nn \\ 
&+& \lambda_{e}\, \bar\ell H e_R+ \lambda_{e}^{\prime}\, \bar\Psi H e_R + \lambda_{N_i}\, \bar \ell \tilde H N_i + \lambda_{N_i}^{\prime}\, \bar\Psi \tilde H N_i  . \label{Lag:leptons}
\eeqn
Here $m_N$ is a Majorana mass matrix for the heavy right-handed neutrinos. To be more general the last term in the Lagrangian should be written as $\lambda_{N_i}^{\prime} \bar\Psi  \tilde H N_i +  \lambda_{N_i}^{\prime \prime} \bar \Psi   \tilde H N^c_i$. We are assuming for simplicity that $\lambda_{N_i}^{\prime} = \lambda_{N_i}^{\prime\prime}$.

The CP asymmetry produced in the annihilation of DM (see Fig.~\ref{Fig:lepasym}) is proportional to the relative phase of the couplings $\lambda_{L_1}$ and $\lambda_{L_2}$ and its calculation is completely analogous to the one for the model of Section~\ref{sec:model}. The relevant washout processes are the same as for the leptogenesis model of Ref.~\cite{Cui:2011ab}.

The decay of $\Psi$ is crucial for the success of the mechanism. 
We explained at the beginning of this section that the $L$-conserving decay $\Psi \to H\,e_R$ has to be subdominant compared to the $L$-violating one, $\Psi \to H\,N$. The corresponding decay widths are
\be
\Gamma_{\Psi\to H e_R} = \frac{\vert \lambda_{e}^{\prime}\vert^2}{32\pi}m_\Psi\,, \qquad \quad \Gamma_{\Psi \to H\,N_i} = \frac{\vert \lambda_{N_i}^{\prime}\vert^2}{16\,\pi}\frac{\left(m_\Psi+m_{N_i}\right)^2\left(m_\Psi^2-m_{N_i}^2\right)}{m_\Psi^3}\,.
\ee
 Asking that $ \Gamma_{\Psi\to H e_R} / \Gamma_{\Psi \to H\,N} \lesssim 0.1$ amounts to the following constraint:
\be \label{eq:decayconstraint}
\vert \lambda_{e}^{\prime}\vert \lesssim 0.2\times \left\vert \lambda_{N_i}^{\prime} \right\vert\times \left(1+ \frac{m_{N_i}}{m_\Psi}\right)\sqrt{1- \frac{m_{N_i}^2}{m_\Psi^2}} .
\ee


\subsection{Numerical results}

\begin{table}[!t]
\centering
\begin{tabular}{ l|cccc|cccc l }
\toprule
Parameter & $m_{\chi}$ & $m_{\Psi}$&  $m_{S_1}$& $m_{S_2}$  & $\lambda_{\chi_1}$ & $\lambda_{\chi_2}$ & $\lambda_{L_1}$ & $\lambda_{L_2}$\\
\midrule
Benchmark & 2 & 2.5  & 6 & 6.6  & 0.95 & 1 & 0.7 & 0.85\\
\bottomrule
\end{tabular}
\caption{Benchmark point for the model with $B-L$ violation. The masses are in TeV. The CP-violating phase $\alpha$, defined by $\lambda_{L_1}\lambda_{L_2}^* \equiv |\lambda_{L_1}| |\lambda_{L_2}| e^{i\alpha}$, is fixed again to $2 \pi / 3$ to maximize the asymmetry.} \label{Tab:benchlep}
\end{table}

We solve the BEs written in Appendix~\ref{app:BEs} with the parameters set to the values in Table~\ref{Tab:benchlep}. We assume that the couplings $\lambda_{e}^{\prime}$ and $\lambda_{N_i}^{\prime}$ satisfy the inequality of Eq.~\eqref{eq:decayconstraint} and that the $L$-violating decay of $\Psi$ is fast enough to keep it in thermal equilibrium.

\begin{figure}[!h]
\begin{center}
\includegraphics[width=0.8\textwidth]{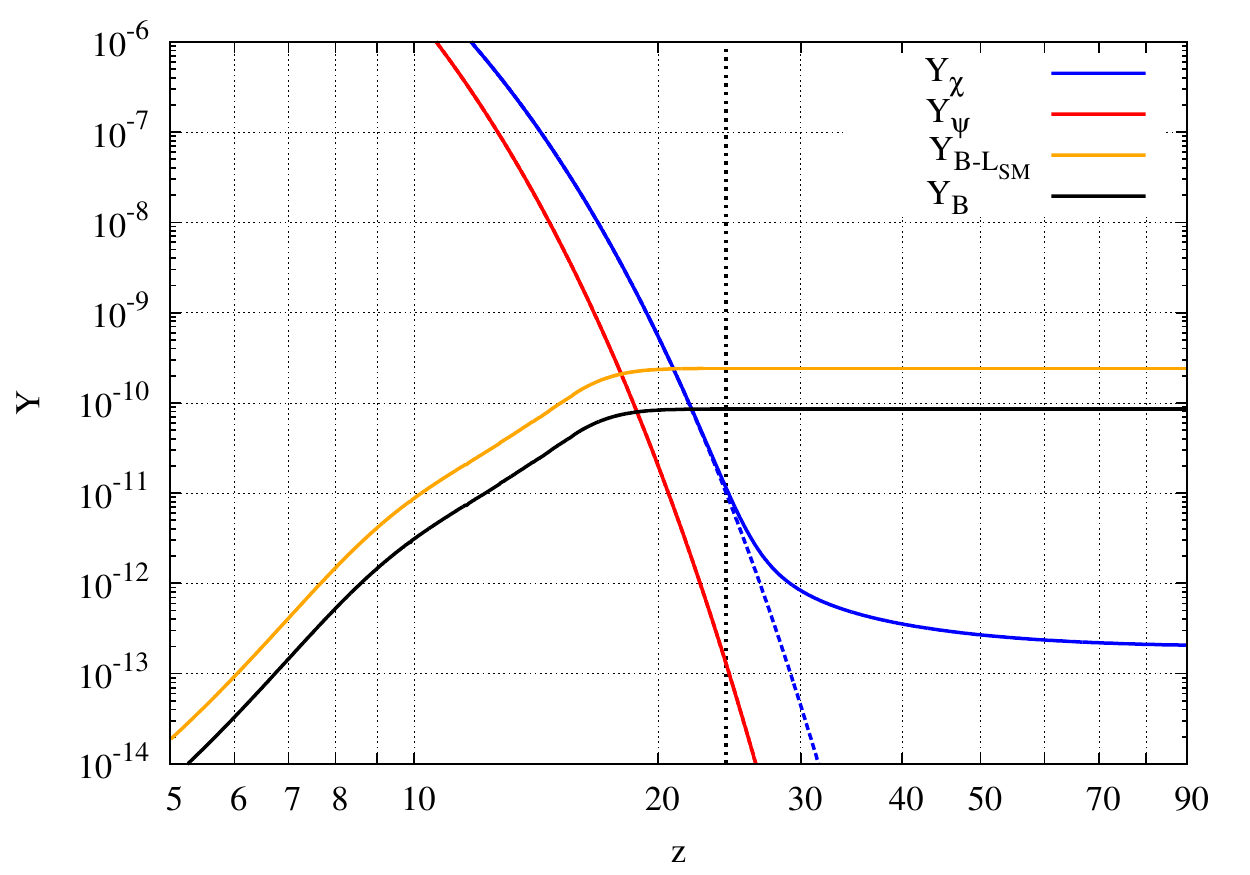}
\end{center}
\vspace{-0.8cm}
\caption{\sl \textbf{\textit{Benchmark for the $B-L$ violating model}  }  We show $Y = n/s$ as a function of $z=m_\chi/T$ for the DM, the exotic field $\Psi$, the asymmetries $Y_{B-L_{\rm SM}}$ and $Y_B$. The blue dashed line traces the DM equilibrium distribution. The vertical black dotted line denotes the approximate point at which sphalerons freeze out. The BEs to get these curves can be found in Appendix~\ref{app:BEs}. The parameters used for this benchmark point are listed in Table~\ref{Tab:benchlep}.
}
\label{Fig:Nico1C}
\end{figure}

We show the evolution of the densities in Fig.~\ref{Fig:Nico1C}. For $z <z_{\rm sfo}$ the baryon asymmetry is related to $Y_{B-L_{\rm SM}}$ via the relation $Y_B = \frac{28}{79} Y_{B-L_{\rm SM}}$ [see Eq.~\eqref{eq:eqviol}], and it remains constant for $z>z_{\rm sfo}$, because in this model $B$ is only violated by the sphaleron processes.

\subsection{Experimental constraints and neutrino masses} \label{neutrinomasses}
The Lagrangian of Eq.~\eqref{Lag:leptons} implies, after electroweak symmetry breaking, the presence of extra massive states: a heavy charged lepton and $2+n$ heavy neutrinos. Assuming  $m_{N_i}\lesssim m_\Psi \sim O$(TeV) , the heavy-particle masses are approximately 
\beqn
\hat{m}_{\Psi^{\pm}}&\simeq & m_\Psi\left(1+\frac{1}{2}\frac{\lambda_{e}^{\prime 2}\,v^2}{2\,m_\Psi^2}\right)\,,
\quad \hat{m}_{N_i}\simeq  m_{N_i} -\frac{ \lambda_{N_i}^{\prime 2}\,v^2}{m_{N_i}}\left(1+\frac{m_{N_i}}{m_\Psi}\right)\,, \nn \\
\hat{m}_{\Psi^{0}_{1}}&= & m_{\Psi}\,,\quad{\rm and}\quad \hat{m}_{\Psi^{0}_{2}}\simeq  m_\Psi+\sum_{i}\frac{\lambda_{N_i}^{\prime 2}\,v^2}{m_{\Psi}}\left(1+\frac{m_{N_i}}{m_\Psi}\right).
\eeqn
Several constraints apply on vector-like leptons. The decays $\Psi^{\pm} \to W^{\pm} \, \Psi^{0} \to $ SM + missing energy were studied at LEP and lead to $\hat{m}_{\Psi^{\pm}}\geq 101.9\,(63.5)$ GeV for  $\hat{m}_{\Psi^{\pm}}-\hat{m}_{\Psi^{0}}\geq 15\,(7)$ GeV~\cite{Beringer:1900zz}. Vector-like leptons manifest themselves in gauge boson self-energies, and electroweak precision data constrain the deviations of the oblique parameter $S$, $T$ or $U$. Ref.~\cite{Arina:2012aj} showed that the $T$ parameter is strongly dependent on the mass splitting between the charge and neutral component of $\Psi$, imposing $ \vert \hat{m}_{\Psi^{\pm}}-\hat{m}_{\Psi^{0}}\vert \lesssim 65$ GeV at $3 \sigma$. This bound in turn implies $\lambda_{N}^\prime (\lambda_{e}^\prime) \lesssim 2 (1) \sqrt{ m_\Psi / 1\,{\rm TeV}}$.
Slightly more stringent constraints are derived in Refs.~\cite{Kannike:2011ng, delAguila:2008pw} from precision data: $\lambda_{e}^\prime\,\lesssim 0.2\times (m_\Psi/1 {\rm TeV})$ and  $\lambda_{N}^\prime\,\lesssim 0.17\times (m_N/1 {\rm TeV})$. Similar limits are derived from inspection of the Higgs decays width~\cite{Moreau:2012da}; in particular, the presence of the vector-like lepton in the loop can explain the possible enhancement of the diphoton channel $H\to \gamma\,\gamma$.

The most stringent constraints stem from lepton flavour violation, as one can expect.
Following Ref.~\cite{Ishiwata:2013gma}, the strongest constraints come from the decays $\mu\to3e$:
\beqn
Br\left(\mu\to3e\right)\simeq 1.7\times 10^{-4} \big\vert \lambda_{e}^\prime\, \lambda_{\mu}^\prime\big\vert^2 \left(\frac{1 {\rm TeV}}{m_{\Psi}}\right)^4\,,
\eeqn
From the experimental limit $Br\left(\mu\to3e\right)\lesssim 1\times 10^{-12}$~\cite{Beringer:1900zz}, assuming equal couplings of $\Psi$ to leptons $\lambda_{e}^\prime=\lambda_{\mu}^\prime=\lambda_{\tau}^\prime=\lambda_{\ell}^\prime$, one obtains
\beqn
\lambda_{\ell}^\prime\lesssim 8.5\times 10^{-3}\left(\frac{m_\Psi}{1 {\rm TeV}}\right)\,.
\eeqn
These constraints are all met in the illustrative benchmark point we used in Table~\ref{Tab:benchlep}.

It is interesting that this $B-L$ violating model requires the presence of heavy neutral Majorana neutrinos to generate the BAU. It turns out that light neutrino masses can then be explained via a TeV-scale seesaw mechanism. However, given that $\Psi$ is an $SU(2)$ doublet, the scenario contemplated here does not differ from the usual TeV-scale seesaw, and at least 2 Majorana $N$'s are mandatory to comply with neutrino oscillations. For vanilla seesaw, the Dirac mass term  $\propto \lambda_{N}\,v\,\overline{\nu_L}\,N$ then has to be suppressed: $\lambda_{N} \simeq 10^{-6} \sqrt{ m_N/ 1{\rm TeV} }$. Note that this suppression ensures that the washouts mediated by $N$ will be negligible.
 Assuming the 2 heavy neutrinos form a pseudo-Dirac pair allows to increase the neutrino Yukawa couplings $\lambda_{N}$, cf. e.g. Ref.~\cite{Ibarra:2010xw}.

Neutrino masses can in principle be accommodated also in the $B-L$ conserving model (see e.g. Ref.~\cite{Branco:1988ex}) discussed in Section~\ref{sec:model}, at the expense of introducing new fields. One could, for instance, implement the case of light Dirac neutrinos~\cite{Bilenky:1987ty}, by adding at least two singlet neutrinos that couple to the active ones via suppressed Yukawa couplings.
Alternatively, one can afford some $L$ violation, if we insist on having Majorana neutrino masses in that model, as long as the $L$ violating processes are out of equilibrium at temperatures close to the sphaleron freeze-out. This way the model would not be $B-L$ conserving strictly speaking, but it would be to a good approximation at the scale where the BarDaMA mechanism is operative. 


\section{Summary}

We have studied the conditions to generate the BAU from the DM annihilations that set the DM relic density. The first thing to note is that this scenario requires a low-scale thermal baryogenesis mechanism. This is because unitarity arguments yield an upper bound of $\sim 340$~TeV for the DM mass, and moreover most WIMP models have a DM mass well below $\sim 10$~TeV. Consequently, we have discussed different known ways to achieve baryogenesis at low temperatures, outlining the particular restrictions that appear when the CP violating interaction is also the one that yields the relic DM abundance. The most attractive mechanism, and the only one that has been proven to work so far, is to have the DM, $\chi$, annihilate into a SM lepton or baryon plus a heavy exotic particle $\Psi$. This way, the dangerous washouts are Boltzmann suppressed while the CP asymmetry can be kept large.

In the CP-violating annihilations a baryon (or lepton) asymmetry is generated in the SM fields, together with an asymmetry of the opposite sign in $\Psi$. If $\Psi$ decayed directly into only SM particles, the baryon asymmetry would get cancelled (in Sec.~\ref{sec:conditions} we explained why this point is actually not so trivial). As argued in~\cite{Cui:2011ab}, for models with a conserved $B-L$ this seems to demand the existence of light sterile particles and a $\mathbb{Z}_n$ discrete symmetry ($n \ge 4$), so that the asymmetry in $\Psi$ is only transferred to and sequestered in this dark sector. Therefore, at present time there would be a matter-antimatter asymmetry in the visible universe and another of opposite sign in the dark sector. We have shown that these requirements of a light dark sector and a $\mathbb{Z}_4$ symmetry are actually not necessary. A general explanation has been provided in Sec.~\ref{sec:conditions} and a concrete model in Sec.~\ref{sec:model}. 

In Sec.~\ref{sec:modelC} we proposed another model that yields the BAU from DM annihilation without a light dark sector, but involving the heavy singlet Majorana neutrinos of the -low scale- type I seesaw, so that in this case $B-L$ is violated. This is interesting because all the models of BarDaMA presented before this work, even those with violation of B-L, contained light sterile degrees of freedom and demanded a $\mathbb{Z}_4$ symmetry. In addition, the heavy Majorana neutrinos play two different roles in the model we presented: they avoid that the asymmetry stored in $\Psi$ cancels the asymmetry in the SM fields and they induce tiny neutrino masses via the seesaw mechanism.

\acknowledgments
We would like to thank Nuria Rius for many interesting discussions and Miguel Nebot for clarifying issues on bounds for vector-like quarks. LU thanks the Center for Theoretical Underground
Physics and Related Areas (CETUP* 2013) in South Dakota for its support and hospitality while part of this work was completed.\\
NB and LU are supported by the DFG TRR33 ``The Dark Universe''.
The work of FXJM is supported by Portuguese national funds through
FCT - Fundação para a Ci\^encia e Tecnologia, project PEst-OE/FIS/UI0777/2011, and by PTDC/FIS-NUC/0548/2012 and CERN/FP/123580/2011 projects. The work of JR has been supported by the Spanish MINECO Subprogramme Juan de la Cierva and it has also been partially supported by the Spanish MINECO grants FPA2011-29678-C02-01, and 
Consolider-Ingenio CUP (CSD2008-00037). In addition JR acknowledges partial support from the  European Union FP7  ITN INVISIBLES (Marie Curie Actions, PITN- GA-2011- 289442).  

\appendix

\section{Cross sections and rates} \label{app:crossNico}
\subsection*{Useful definitions}
We assume Maxwell-Boltzmann statistics for the various species $X$, 
\beqn
n^{eq}_{X}(m_X,T)=\frac{g_X\,m_X^2\,T}{2\pi^2}\,K_2\left(\frac{m_X}{T}\right)\,.
\eeqn
Note that for massless species, $n^{eq}_{X}(0,T)= g_X T^3 / \pi^2$.
$g_X$ is the numbers of degrees of freedom for the particle $X$: $g_{S_\alpha}=1$, $g_{U_R}=3$, $g_{Q}=6$, $g_{N_i}=g_\ell=g_{H}=g_\chi=2$, while $g_{\Psi}=2\,N_c$ $=6(4)$ in the $B-L$ conserving (violating) scenario considered.  
The entropy density and the Hubble expansion rate in a radiation dominated universe are
\beqn
s(T)=g_*\,\frac{2\pi^2}{45}\,T^3\,,\quad H(T)=\sqrt{\frac{4\pi^3\,g_*}{45}}\frac{T^2}{m_{pl}}\,
\eeqn
with the Planck mass $m_{pl}\simeq 1.22\times 10^{19}$ GeV, and $g_*$ the effective number of relativistic degrees of freedom, that we took constant and equal to the SM value $g_*=106.75$.

\subsection*{Scattering rate densities}

For the process $i+j\to k+l$, the interaction rate density $\gamma$ is defined by:
\begin{equation} \label{eq:rate}
\gamma(z)=\frac{m_{\chi}^4}{64\,\pi^4\,z} \int_{x_{min}}^{\infty}dx \sqrt{x}\,\sigma_R(x\,m_\chi^2)\,K_1\left(z\,\sqrt{x}\right)\,,
\end{equation}
where $z\equiv\mx/T$, $x_{min}={\rm Max}\left\lbrace\left(\frac{m_i+m_j}{m_\chi}\right)^2,\,\left(\frac{m_k+m_l}{m_\chi}\right)^2\right\rbrace$. Note that $x \, m^2_\chi = s$, with $s, t ,u$ the Mandelstam variables\footnote{We use the symbol $s$ both for the entropy density and for the Mandelstam variable. However it is always clear from the context which quantity we are referring to.}.
The reduced cross section $\sigma_R$ is related to the total cross section $\sigma$ via
\begin{equation}
\sigma_R(s)=\frac{2\,\lambda(s,m_i^2,m_j^2)}{s}\sigma(s)\,,\quad \lambda(s,m_i^2,m_j^2)\equiv \left(s-(m_i+m_j)^2\right)\left(s-(m_i-m_j)^2\right).
\end{equation}
It is given by
\begin{equation}
\sigma_R(s)\equiv\frac{1}{8\pi\,s}\int_{t_0}^{t_1}|M|^2\,dt\,,
\end{equation}
where $|M|^2$ is summed (not averaged) over all the degrees of freedom of the initial and final states, and the integration limits are 

\begin{equation*}
t_{0(1)}=\frac{1}{4\,s}\left(m_i^2-m_j^2-m_k^2+m_l^2\right)^2-\frac{1}{4\,s}\left(\sqrt{\lambda(s,m_i^2,m_j^2)}\mp\sqrt{\lambda(s,m_k^2,m_l^2)}\right)^2\,.
\end{equation*}

\subsection*{Decay rate densities}

\noindent
For the decays $i\to n+m$, the interaction rate density is defined by:
\begin{equation}
\gamma(z)=n^{eq}_{i}\left(m_i,m_\chi/z\right)\frac{K_1(z\,m_i/m_\chi)}{K_2(z\,m_i/m_\chi)}\,\Gamma\,=\frac{\mx\,m_i^2}{2\,\pi^2\,z}K_1\left(\frac{m_i}{\mx}z\right)\,\Gamma^s\,,
\end{equation}
where $\Gamma$ is the total decay width of the particle $i$,
\beqn
\Gamma(i\to j\,k)= \frac{1}{16\pi}\frac{\sqrt{\lambda(m_i^2,m_j^2,m_k^2)}}{m_i^3}\vert M(i\to j\,k)\vert^2\, ,
\eeqn
while the superscript in $\Gamma^s$ indicates that such a quantity is similar to $\Gamma$ but is obtained by summing, rather than averaging, over the degrees of freedom of the decaying particle.

In the following we present a collection of the squared matrix elements $|M|^2$ for the scattering and decay processes relevant to the $B-L$ conserving model of Section~\ref{sec:model}, for which $N_c = 3$ is the number of colors. The $|M|^2$ for the $B-L$ violating case are obtained from the ones we list below by substituting $U_R \to \ell$, $N_c \to 1$, $\lambda_{B_\alpha} \to \lambda_{L_\alpha}$ and $m_U \to m_\ell$. We include the SM quark (lepton) mass $m_U$ ($m_\ell$) for completeness, although for all practical purposes they can be neglected in actual calculations.\\

The processes we consider are all mediated by the scalar $S_{1,2}$, so for convenience we introduce the propagators 
\beqn
D_{C,i}=\left(C-m_{S_i}^2 \right)\,,\quad C=s,t,u\,.
\eeqn 
Special attention should be paid to the $s-$channel, in which case
the propagators should be written in the Breit-Wigner form
\be \label{eq:dress}
D_{s,\alpha}^{-1}=\left( s-m^2_{S_\alpha} - i\, {\rm Im}\left[\Pi_\alpha (m^2_{S_\alpha}) \right] \right)^{-1},
\ee
where $\Pi_\alpha(s)$ is given by the 1PI (one Particle Irreducible) insertions into the propagator. At one loop $\Pi_\alpha(s)$ can be computed using the Cutkosky rules. We find
\be 
{\rm Im}\left[\Pi_{1,2}(s)\right] = -\frac{\lambda_{\chi_{1,2}}^2}{16 \pi} \sqrt{s(s-4m_\chi^2)} - N_c \frac{\lambda_{B_{1,2}}^2}{8\pi} \frac{(s-m_\Psi^2)^2}{s}\,.\nonumber
\ee
Note that
\be
{\rm Im} \left[\Pi_\alpha(m^2_{S_\alpha})\right] = -  m_{S_\alpha}\Gamma_\alpha \, , \nn
\ee
where $\Gamma_\alpha$ is the total decay width of $S_\alpha$.
Indeed, when $\Pi_\alpha(m^2_{S_\alpha})$ is small, one recovers the more familiar expression
\be \label{eq:BW}
D_{s,\alpha}^{-1}=\left(s-m^2_{S_\alpha} + i  m_{S_\alpha}\Gamma_\alpha \right)^{-1}\,.
\ee
For some values of the parameters the resonance can be broad, in which case one should use Eq.~\eqref{eq:dress} rather than Eq.~\eqref{eq:BW}.
Note that neglecting $\Pi_\alpha$ and using instead an undressed propagator, $(s-m^2_{S_\alpha})^{-1}$, in the Benchmark A example in Section~\ref{sec:model} is numerically not important, as the pole develops at large exchanged momentum $s \sim m_{S_\alpha}^2 \gg 4\,m_\chi^2$, for which an important Boltzmann suppression occurs, owing to the Bessel factor in Eq.~\eqref{eq:rate}.

\subsubsection*{Annihilation}
\subsubsection*{ $\bullet\quad \boldsymbol{\chi\,\chi\to\bar\Psi\, U_R}$}
\beqn \label{XXtp}
\vert M\vert^2= \frac{1}{2}\,N_c\,\times 2\,s\,(s-\mps^2-\mf^2)\times\left\vert \frac{\lxo\,\lbo}{D_{s,1}}+\frac{\lxt\,\lbt}{D_{s,2}}\right\vert^2\,.
\eeqn

\subsubsection*{Washouts}
\subsubsection*{$\bullet\quad \boldsymbol{\chi\, U_R\to\chi\,\Psi}$}
The process $\chi\, U_R\to\chi\,\Psi$ is derived from $\chi\,\chi\to\bar\Psi\, U_R$ by crossing symmetry, replacing $s\to t$ in Eq.~\eqref{XXtp}.

\subsubsection*{$\bullet\quad \boldsymbol{\bar U_R\,\Psi\to\bar\Psi\, U_R}$}
\beqn \label{UPPU}
&&\vert M\vert^2= \left\vert \frac{\lbo^2}{D_{s,1}}+\frac{\lbt^2}{D_{s,2}}\right\vert^2 \,\left(s-\mps^2-\mf^2\right)^2+\left\vert \frac{\lbo^2}{D_{t,1}}+\frac{\lbt^2}{D_{t,2}}\right\vert^2 \,\left(t-\mps^2-\mf^2\right)^2\nonumber \\
&&+ 2{\rm Re}\left[ \left(\frac{\lbo^2}{D_{s,1}}+\frac{\lbt^2}{D_{s,2}}\right)\left(\frac{\lbo^{2\,*}}{D_{t,1}}+\frac{\lbt^{2\,*}}{D_{t,2}}\right)\right]\,\left(s\,t-\mps^4-\mf^4\right).
\eeqn

\subsubsection*{$\bullet\quad \boldsymbol{U_R\, U_R\to\Psi\,\Psi}$}
The squared matrix element for $U_R\, U_R\to\Psi\,\Psi$ is again obtained by crossing symmetry, replacing in Eq.~ \eqref{UPPU} $s\to t$ and $t\to u$.

\subsubsection*{Decays}
\subsubsection*{$\bullet\quad\boldsymbol{\Psi\to H\,Q}$}
\begin{eqnarray}
&|M|^2& = \lp^2\left(\mps^2+\mq^2-\mh^2\right)\,,\nonumber \\
&\Gamma^s & = \frac{\lp^2\,N_c}{8\pi}\frac{\sqrt{\lambda(\mps^2,\mq^2,\mh^2)}}{\mps^3}\left(\mps^2+\mq^2-\mh^2\right)
\end{eqnarray}

\subsubsection*{$\bullet\quad\boldsymbol{S_1\to\chi\,\chi}$}

\begin{eqnarray}
&|M|^2& =2\,\lxo^2\,\mso^2\,,\nonumber \\
&\Gamma^s & =\frac{1}{2}\frac{\lxo^2}{8\pi}\sqrt{\mso^2-4\,\mx^2}\,.
\end{eqnarray}
The factor $1/2$ takes into account the identical final state particles.

\subsubsection*{$\bullet\quad\boldsymbol{S_1\to\bar\Psi\,U_R $ + $S_1\to\bar U_R\,\Psi}$}
\begin{eqnarray}
&|M|^2& = N_c\,|\lbo|^2(\mso^2-\mf^2-\mps^2)\,,\nonumber \\
&\Gamma^s & = 2\,N_c\,\frac{|\lbo|^2}{16\pi}(\mso^2-\mf^2-\mps^2)\frac{\sqrt{\lambda(\mso^2,\mps^2,\mf^2)}}{\mso^3}\,,
\end{eqnarray}
and analogously for the decays of $S_2$.

\subsection*{The CP asymmetry in scatterings}
We introduce for convenience the functions
\beqn
&&f_{S}(m_{S_\alpha})=s\,(s-m_{S_\alpha}^2)(m_{S_\alpha}^2s-\mps^4-\mf^4)\log\left[s\frac{s-2\mps^2-2\mf^2+m_{S_\alpha}^2}{m_{S_\alpha}^2s-(\mps^2-\mf^2)^2}\right]\,, \nonumber \\
&&f_{V}(m_{S_\alpha})=-\left[(\mps^2+\mf^2)^2-(2\mps^2+m_{S_\alpha}^2+2\mf^2)\,s+2s^2\right]
\left[\mps^4+(s-\mf^2)^2-2\mps^2\,(s+\mf^2)\right]\,.\nonumber 
\eeqn
The function $f_{S (V)}$ originates from the cut through the loop of the self-energy (vertex) diagram (see e.g. Fig.~\ref{Fig:CPandwashout}). The rate for the asymmetry $\Delta\gamma(\chi\chi\to\bar\Psi U_R)$ is then obtained, after integration, from the reduced cross section
\begin{eqnarray}
& &\Delta\sigma_R\left(\chi\chi\to\bar\Psi\,U_R\right)=\frac{N_c}{8 \pi^2}\frac{\sqrt{s-4\,m_{\chi}^2}}{s^{3/2}\,(s-\mso^2)(s-\mst^2)}\times  \nonumber \\
& &\Bigg\lbrace 2 \lxo\lxt\,\text{Im}\left(\lbo\,\lbt^*\right) \left( |\lbo|^2 \frac{f_{S}(\mso)+f_{V}(\mso)}{s-\mso^2}-|\lbt|^2 \frac{f_{S}(\mst)+f_{V}(\mst)}{s-\mst^2}\right)  \nonumber \\
&&-\text{Im}\left(\lbo^2\,\lbt^{*2}\right)\left( \lxo^2 \frac{f_{S}(\mst)+f_{V}(\mst)}{s-\mso^2}
-\lxt^2 \frac{f_{S}(\mso)+f_{V}(\mso)}{s-\mst^2} \right) \Bigg\rbrace\,.
\end{eqnarray}


\section{Derivation of the Boltzmann equations} \label{app:BEs}
The BEs for thermal baryogenesis models have been derived in many works (e.g. see~\cite{Giudice:2003jh} and~\cite{Nardi:2007jp} for many details on different issues). Therefore we will simply state most of the results. However, there is a particular feature in our model and those of ~\cite{Cui:2011ab,Bernal:2012gv} which deserves special attention, namely that the CP asymmetry is not only generated in the decay of heavy scalars $S_\alpha \, \{\alpha=1,2\}$, but also in the annihilations mediated by them. More specifically, we want to show how to do the on shell subtractions of the CP-violating rates to get -classical- BEs that respect unitarity (i.e. that no asymmetry is generated in equilibrium). 

We take as an example the model introduced in Sec.~\ref{sec:model} and  derive the BEs for the $B_{SM} - L$ asymmetry, $Y_{B_{SM} - L}$, where $B_{SM}$ represents the baryon number in the SM fields. As usual, it is convenient to choose $B_{SM} - L$ as the asymmetry to evolve, since it is not affected by the sphalerons. For simplicity we assume that $m_{S_2} \gg m_{S_1}$, so that only $S_1$ can be produced on-shell during the relevant epoch of baryogenesis\footnote{This is just to avoid clutter. Relaxing the assumption $m_{S_2} \gg m_{S_1}$ simply results in substituting $S_1 \to \sum_\alpha S_\alpha$ in the following formulae.}, and also that $\Psi_j \, \{j=1,2\}$ (with $\Psi_1 \equiv \Psi$) couple to just one flavour of SM quarks, $U_R$ and $Q_L$. The contribution of the relevant CP-violating interactions to the evolution of $Y_{B_{SM} - L}$ reads
\begin{equation}
\label{eq:be1}
\begin{split}
3 s z H \frac{\dif Y_{B_{SM}-L}}{\dif z} & = \ydyeq{S_1} \ghor{S_1}{\bar \Psi U_R} - \ydyeq{\bar \Psi} \ydyeq{U_R} \ghor{\bar \Psi U_R}{S_1} \\
& -  \ydyeq{S_1} \ghor{S_1}{\Psi \bar U_R} + \ydyeq{\Psi} \ydyeq{\bar U_R} \ghor{\Psi \bar U_R}{S_1} \\
& + 2 \ydyeq{\bar U_R} \ydyeq{\Psi} \gphor{\bar U_R \Psi}{\bar \Psi U_R} - 2 \ydyeq{\bar \Psi} \ydyeq{U_R} \gphor{\bar \Psi U_R}{\Psi \bar U_R} \\
& + \ydyeqs{\chi} \gphor{\chi \chi}{\bar \Psi U_R} - \ydyeq{\bar \Psi} \ydyeq{U_R} \gphor{\bar \Psi U_R}{\chi \chi} \\ 
& - \ydyeqs{\chi} \gphor{\chi \chi}{\Psi \bar U_R} + \ydyeq{\Psi} \ydyeq{\bar U_R} \gphor{\Psi \bar U_R}{\chi \chi} \\
& + \text{other scatterings} + \text{decay of $\Psi_j$} \;.
\end{split}
\end{equation}
The factor 3 in the left hand side arises because $U_R$ has baryon number 1/3 and the factor 2 in the right hand side goes when a process violates $B$ by $ 2 \times 1/3$ units. The ``other scatterings'' and ``decay of $\Psi_j$'' terms will be specified at the end, since their inclusion is quite trivial. 

The crucial point is that, in order to avoid double counting, it has been necessary to introduce the $\gamma'$ rates in Eq.~\eqref{eq:be1}. These are the total rates with the on-shell part subtracted, i.e. they only involve off-shell contributions (see~\cite{kolb79} or~\cite{Giudice:2003jh} for the specific case of leptogenesis):
\begin{equation}
\label{eq:sub1}
\begin{split}
\gphor{\bar U_R \Psi}{\bar \Psi U_R} & = \ghor{\bar U_R \Psi}{\bar \Psi U_R} - \ghor{\bar U_R \Psi}{S_1} \mibr{S_1}{\bar \Psi U_R} \; ,\\
\gphor{\chi \chi}{\bar \Psi U_R} & = \ghor{\chi \chi}{\bar \Psi U_R} - \ghor{\chi \chi}{S_1} \mibr{S_1}{\bar \Psi U_R} \;,
\end{split}
\end{equation}
with similar expressions holding for the remaining rates.

To first order in the CP asymmetries 
\beqn
\frac{\dghor{a,b,\dots}{c,d,\dots}}{\ghor{a,b,\dots}{c,d,\dots}+\ghor{\bar a,\bar b,\dots}{\bar c,\bar d,\dots}} \equiv \frac{\ghor{a,b,\dots}{c,d,\dots}-\ghor{\bar a,\bar b,\dots}{\bar c,\bar d,\dots}}{\ghor{a,b,\dots}{c,d,\dots}+\ghor{\bar a,\bar b,\dots}{\bar c,\bar d,\dots}}\,,\nonumber
\eeqn and using CPT, Eq.~\eqref{eq:be1} becomes
\begin{equation}
\label{eq:be2}
\begin{split}
3 s z H \frac{\dif Y_{B_{SM}-L}}{\dif z} & = \left(\ydyeq{S_1} + \ydyeq{\Psi + \bar \Psi}\right) \dghor{S_1}{\bar \Psi U_R} + 2 \ydyeq{\Psi + \bar \Psi} \dgphor{\bar U_R \Psi}{\bar \Psi U_R} \\
& + \left(\ydyeqs{\chi} + \ydyeq{\Psi + \bar \Psi} \right) \dgphor{\chi \chi}{\bar \Psi U_R} \\
& -\left(\ydyeq{\Psi + \bar \Psi} y_{U_R} - y_\Psi \right) \left[ \ghor{S_1}{\bar \Psi U_R} + 2 \gphor{\bar U_R \Psi}{\bar \Psi U_R} \right. \\ & \left. + \gphor{\chi \chi}{\bar \Psi U_R}\right] + \text{other scatterings} + \text{decay of $\Psi_j$} \; ,
\end{split}
\end{equation}
where we have defined $y_{X} = \frac{Y_{X} - Y_{\bar X}}{Y_X^{eq}}$ and $Y_{X + \bar X} = Y_X + Y_{\bar X}$ for any particle $X$. 

Unitarity and CPT imply that
\begin{equation}
\begin{split}
& \ghor{U_R \bar \Psi}{U_R \bar \Psi} + \ghor{U_R \bar \Psi}{\bar U_R \Psi} + \ghor{U_R \bar \Psi}{\chi \chi} = \\
& \ghor{\bar U_R \Psi}{\bar U_R \Psi} + \ghor{\bar U_R \Psi}{U_R \bar \Psi} + \ghor{\bar U_R \Psi}{\chi \chi} \; ,
\end{split}
\end{equation} 
where $S_1$ has been considered as an intermediary resonance and not an actual ``out'' state. Therefore,
\begin{equation}
\dghor{U_R \bar \Psi}{\bar U_R \Psi} + \dghor{U_R \bar \Psi}{\chi \chi} = 0 \;.
\end{equation}
It follows that at first order in the CP asymmetries
\begin{equation}
\begin{split}
& \dgphor{\chi \chi}{U_R \bar \Psi} = \dghor{\chi \chi}{U_R \bar \Psi} - \dghor{\chi \chi}{S_1} \mibr{S_1}{U_R \bar \Psi} \\ & \hspace{3.4cm} - \mibr{S_1}{\chi \chi} \dghor{S_1}{U_R \bar \Psi} \\
& \; = \dghor{\chi \chi}{U_R \bar \Psi} - \mibr{S_1}{\chi \chi} \dghor{S_1}{U_R \bar \Psi} \; ,\\
& \dgphor{\bar U_R \Psi}{U_R \bar \Psi} = \dghor{\bar U_R \Psi}{U_R \bar \Psi} - \dghor {\bar U_R \Psi}{S_1} \mibr{S_1}{U_R \bar \Psi} \\ & \hspace{3.7cm} - \mibr{S_1}{\bar U_R \Psi} \dghor{S_1}{U_R \bar \Psi} \\
& \; = \dghor{\bar U_R \Psi}{U_R \bar \Psi} - \dghor{S_1}{U_R \bar \Psi} \left[ \mibr{S_1}{U_R \bar \Psi} + \mibr{S_1}{\bar U_R \Psi} \right] \; .
\end{split}
\end{equation} 
Using these results in Eq.~\eqref{eq:be2} and writing the explicit expressions for the  ``other scatterings'' and ``decay of $\Psi_j$'' terms, we arrive at  a BE that clearly respects all the Sakharov conditions~\cite{Sakharov:1967dj}:
\begin{equation}
\begin{split}
\label{eq:be3}
& 3 s z H \frac{\dif Y_{B_{SM}-L}}{\dif z} = \left(\ydyeqs{\chi} - \ydyeq{\Psi + \bar \Psi} \right) \dghor{\chi \chi}{\bar \Psi U_R} \\& \quad + \left[ \left(\ydyeq{S_1} - \ydyeq{\Psi + \bar \Psi}\right)  - \mibr{S_1}{\chi \chi}  \left(\ydyeqs{\chi} - \ydyeq{\Psi + \bar \Psi} \right) \right] \dghor{S_1}{\bar \Psi U_R}\\
& \quad -\left(\ydyeq{\Psi + \bar \Psi} y_{U_R} - y_\Psi \right) \Big[ \ghor{S_1}{\bar \Psi U_R} + 2 \gphor{\bar U_R \Psi}{\bar \Psi U_R} + \gphor{\chi \chi}{\bar \Psi U_R} \Big] \\ 
& \quad - 4 \left(y_{U_R} - \ydyeq{\Psi + \bar \Psi} y_\Psi \right) \ghor{U_R U_R}{\Psi \Psi} - \left(\ydyeq{\chi} y_{U_R} -  \ydyeq{\chi} y_\Psi \right) \ghor{\chi U_R}{\chi \Psi} \\
& \quad - \sum_{j=1,2} \left(y_{Q_L} + y_H - y_{\Psi_j} \right) \ghor{\Psi_j}{H Q_L}  \;.
\end{split}
\end{equation}
The asymmetry $\dghor{\chi \chi}{\bar \Psi U_R}$ is $\order{\lambda_B^4 \lambda_X^2}$ and does not decay exponentially for $T \lesssim m_{S_1}$, hence it is the one that ``survives'' in the effective approach of~\cite{Bernal:2012gv} and it is also the only one we consider in the numerical results of this work. Let us remark that due to unitarity and CPT, $\dghor{\chi \chi}{\bar \Psi U_R} = \dghor{U_R \bar \Psi}{\bar U_R \Psi}$, with the lowest order contribution to this last asymmetry coming from a loop involving the DM $\chi$ (see Fig.~\ref{Fig:CPTunit}).
Furthermore, the equality
\begin{equation*}
\begin{split}
& \ghor{S_1}{\bar \Psi U_R} + 2 \gphor{\bar U_R \Psi}{\bar \Psi U_R} + \gphor{\chi \chi}{\bar \Psi U_R} = \\
& 2 \ghor{\bar U_R \Psi}{\bar \Psi U_R} + \ghor{\chi \chi}{\bar \Psi U_R} \; ,
\end{split}
\end{equation*}
valid at lowest order, allows to write the BE~\ref{eq:be3} entirely in terms of non-primed rates.
Lastly, note that for simplicity we have not included all the terms that violate $B_{SM}-L$, but only the most relevant to our numerical results (see Sec.~\ref{sec:model}).

Under the conditions stated in Sec.~\ref{sec:model}, two more simplifications can be done in Eq.~\eqref{eq:be3}. First, the fast processes $\lrproname{\chi \chi}{\Psi \bar U_R}$ and $\lrproname{\Psi}{S_3 \Psi_2}$ keep $\Psi$ -almost- in thermal equilibrium, so that $Y_{\Psi + \bar \Psi} = Y_{\Psi + \bar \Psi}^{eq}$ (at zeroth order in the CP asymmetries). Second, since the $m_{S_\alpha}$ were taken considerably larger than $m_{\chi}$ and $m_{\Psi}$ (both for simplicity and to have the BAU originated solely from DM annihilations), all the processes involving an on shell $S_{\alpha}$ can be neglected. Hence, an appropriate set of BE for the model in Sec.~\ref{sec:model} is given by 
\begin{eqnarray}
 3 s z H \frac{\dif Y_{B_{SM}-L}}{\dif z} & = & \left(\ydyeqs{\chi} - 1 \right) \dghor{\chi \chi}{\bar \Psi U_R} \notag \\
& & - \left(y_{U_R} - y_\Psi \right) \left[ 2 \ghor{\bar U_R \Psi}{\bar \Psi U_R} + \ghor{\chi \chi}{\bar \Psi U_R}\right] \notag \\
& &  - 4 \left(y_{U_R} - y_\Psi \right) \ghor{U_R U_R}{\Psi \Psi} - \ydyeq{\chi} \left(y_{U_R} -  y_\Psi \right) \ghor{\chi U_R}{\chi \Psi} \notag \\
& & - \sum_{j=1,2} \left(y_{Q_L} + y_H - y_{\Psi_j} \right) \ghor{\Psi_j}{H Q_L} \; , \label{eq:be4}\\ 
s z H \frac{\dif Y_{\chi}}{\dif z} & = & - 4 \left[ \ydyeqs{\chi} - 1 \right] \ghor{\chi \chi}{\bar \Psi U_R} \;. \label{eq:be4b}
\end{eqnarray}

To solve this set of BEs it is necessary to express the density asymmetries $y_{U_R, Q_L, H, \Psi}$ of Eq.~\eqref{eq:be4} in terms of the $B_{SM}-L$ asymmetry, which can be done by considering the conservation laws and the chemical equilibrium conditions due to fast interactions (see e.g.~\cite{harvey90,Khlebnikov:1996vj,nardi05}).
 The scenario considered in this work takes place at temperatures $T \ll 10^5$~GeV, hence all the Yukawa interactions of the SM are in equilibrium. Furthermore, we also take the electroweak sphalerons to be fast, since we are interested in the evolution of the asymmetry before the sphalerons decouple. Finally, the conservation of the hypercharge and of $B_{\Psi} + B_{\Psi_2} + B_{SM} - L$ must also be taken into account, together with the condition $\mu_{\Psi}=\mu_{\Psi_2}$ due to the fast process $\lrproname{\Psi}{S_3 \Psi_2}$. Putting all together we get~\footnote{A comment regarding our derivation is in due order. The equilibrium conditions we impose are valid in the symmetric phase. In this regard our calculations would be valid if the sphalerons decoupled right below the critical temperature, $T_c\sim 140$ GeV, corresponding to a strong phase transition. However, in the SM this transition is smooth, and the sphalerons decouple at a lower temperature, $T_{\rm sfo}\sim 80$ GeV in the SM, their diffusion rate being exponentially suppressed at lower temperatures. Between $T_c$ and $T_{\rm sfo}$, part of the lepton asymmetry is still transferred to the baryon sector, however with varying coefficient. At the end of the transition, as different chemical equilibrium conditions hold, the $B \leftrightarrow B-L$ conversion is modified, however slightly (in the SM, from $Y_{B}=28/79\,Y_{B-L}$ at $T \sim T_c$  to $Y_{B}=12/37\,Y_{B-L}$ at $T \sim T_{\rm sfo}$.). We use in our numerical analysis a sphaleron decoupling temperature $T_{\rm sfo}\sim T_c/1.7$, although we derive the equilibrium conditions in the symmetric phase. Derivation of these conversion factors along the electroweak crossover is beyond the scope of our analysis, and numerically only induces $O(10\%)$ corrections.}
\begin{align}
Y_{\Delta Q_L} &= g_Q\frac{4}{237} Y_{B_{SM}-L}\;, \qquad & Y_{\Delta U_R } &= g_{U_R}\frac{31}{237} Y_{B_{SM}-L}\;, \notag \\
Y_{\Delta H} &= g_H \frac{18}{79} Y_{B_{SM}-L}\;, \qquad & Y_{\Delta \Psi} &= -\frac{f}{1+f} Y_{B_{SM}-L}\;, \notag \\
Y_{\Delta\Psi_2} &= - \frac{1}{1+f} Y_{B_{SM}-L}\;, \qquad & Y_{L} &= - \frac{63}{79} Y_{B_{SM}-L} \;.  \label{eq:Yrel}
\end{align}
with 
\begin{equation*}
f=f(m_\Psi,m_{\Psi_2},T) \equiv \frac{m_\Psi^2 K_2(m_\Psi/T)}{m_{\Psi_2}^2 K_2(m_{\Psi_2}/T)} \simeq \left(\frac{m_\Psi}{m_{\Psi_2}}\right)^{\frac{3}{2}} e^{-(m_\Psi-m_{\Psi_2})/T} \quad (T \ll m_\Psi-m_{\Psi_2}) \; .
\end{equation*}

One final comment is in order. The BEs have been derived assuming kinetic equilibrium and Maxwell-Boltzmann statistics for all the particles. This is usually a good approximation, especially in the strong washout regime~\cite{basboll06,garayoa09,hahnwoernle09}. However, when spectator processes are taken into account, the -correct- use of quantum statistical distributions (including the Fermi-Dirac blocking factor and the stimulated emission factor for bosons) brings a relative factor of 1/2 between the washout terms induced by bosons and fermions (see~\cite{fong10III} or the Appendix A of~\cite{fong11} for details), which is not negligible. The reason is that what really multiplies the rates in the washout part is not the difference between the density asymmetries, but the difference between the corresponding chemical potentials. This effect can be taken into account by replacing $y_X$ by ${\scriptstyle \mathcal Y}_X$ for the massless particles in the above BEs, where ${\scriptstyle \mathcal Y}_X \equiv Y_{X - \bar X}/Y_{f}^{eq}\;$ $( Y_{X - \bar X}/Y_{s}^{eq})$ for fermions (bosons) and $Y_{f}^{eq} \equiv \tfrac{1}{2} Y_{s}^{eq} \equiv \tfrac{15}{8 \pi^2 g_*}$. 

The derivation of the BEs for the model presented in Sec.~\ref{sec:modelC} is very similar, hence we just give the final expressions here. Again we only include those processes that are relevant under the conditions stated in Sec.~\ref{sec:modelC} and we recall that the vector-like lepton $\Psi$ couples only to one flavor of SM leptons, $\ell$. Also note that since the exotic particles do not carry baryon number, $B_{SM}=B$.
\begin{eqnarray}
 - s z H \frac{\dif Y_{B-L_{SM}}}{\dif z} & = & \left(\ydyeqs{\chi} - 1 \right) \dghor{\chi \chi}{\bar \Psi \ell} \notag \\
& & - \left(y_{\ell} - y_\Psi \right) \left[ 2 \ghor{\bar \ell \Psi}{\bar \Psi \ell} + \ghor{\chi \chi}{\bar \Psi \ell}\right] \notag \\
& &  - 4 \left(y_{\ell} - y_\Psi \right) \ghor{\ell \ell}{\Psi \Psi} - \ydyeq{\chi} \left(y_{\ell} -  y_\Psi \right) \ghor{\chi \ell}{\chi \Psi} \; , \label{eq:beC}\\ 
s z H \frac{\dif Y_{\chi}}{\dif z} & = & - 4 \left[ \ydyeqs{\chi} - 1 \right] \ghor{\chi \chi}{\bar \Psi \ell} \;. \label{eq:beCb}
\end{eqnarray}

These BEs are complemented by a set of conservation laws and chemical equilibrium conditions. In particular, for this model there is not a $B-L$ conserved charge, but the fast process $\lrproname{\Psi}{\bar H N}$ involving the heavy Majorana singlet $N$ implies that $\mu_\Psi = - \mu_H$. For $T \ll m_\Psi$ we get the following relations,
\begin{align}
Y_{\Delta\ell} &= -g_{\ell}\frac{221}{711} Y_{B-L_{SM}}\;, \qquad & Y_{\Delta H} &= -g_{H}\frac{8}{79} Y_{B-L_{SM}}\;, \notag \\
Y_{\Delta \Psi } &= -\frac{8}{79} g Y_{B-L_{SM}}\;, \qquad & Y_{B} &= \frac{28}{79} Y_{B-L_{SM}} \;, \label{eq:eqviol}
\end{align}
with 
\begin{equation*}
g=g(m_\Psi,T) \equiv -\frac{6}{(2\pi)^{3/2}} \frac{g_{\Psi}}{g_{H}} \left(\frac{m_{\Psi}}{T}\right)^{\frac{3}{2}} e^{-m_\Psi/T} \; .
\end{equation*}
Recall that here $g_{\Psi}=4$.


\subsection*{An explicit verification of CPT and unitarity}
\begin{figure}[!t]
\begin{center}
\includegraphics[width=1.0\textwidth]{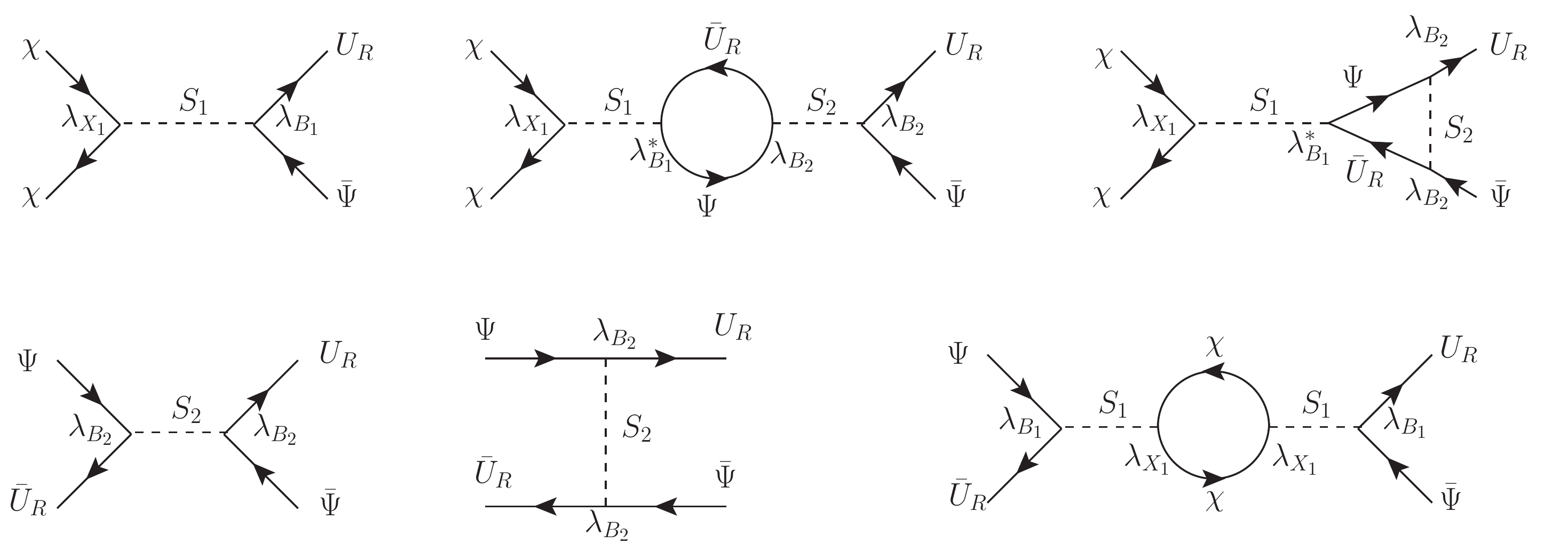}
\end{center}
\vspace{-0.8cm}
\caption{\sl \textbf{\textit{Diagrams for an explicit verification of CPT and unitarity.}} }
\label{Fig:CPTunit}
\end{figure}

In the derivation of the BEs we stated that unitarity and CPT imply the following relation for the rates:
\be \label{eq:CPT}
\Delta \gamma (\bar U_R \Psi \to U_R \bar\Psi ) = - \Delta\gamma (\chi \chi \to U_R \bar\Psi ).
\ee
In this subsection we show that this can be verified explicitly at one loop. One has to compute first the squared matrix elements, then the reduced cross sections that enter the definition of the rates in the equation above. We have already listed the result for the reduced cross section corresponding to the asymmetry in $\chi \chi \to U_R \bar\Psi (\bar U_R \Psi)$. It is the sum of various terms, each of which is proportional to a different combination of the couplings $\lambda$'s. The relation of Eq.~\eqref{eq:CPT} has to hold for each such combination separately. As an example, let us pick the one shown in the diagrams of Fig.~\ref{Fig:CPTunit}. The reduced cross section coming from the diagrams at the top of Fig.~\ref{Fig:CPTunit} is
\beqn 
&&-\lxo^2\,\text{Im}\left(\lbo^2\,\lbt^{*2}\right)\frac{N_c\,\sqrt{s-4\mx^2}}{8\pi^2\,s^{3/2}\,(s-\mso^2)^2\,(s-\mst^2)}\nonumber\\
&&\times\Bigg\{-\left[(\mps^2+\mf^2)^2-(2\mps^2+\mst^2+2\mf^2)\,s+2s^2\right]\left[\mps^4+(s-\mf^2)^2-2\mps^2(s+\mf^2)\right]\nonumber\\
&&\qquad+s\,(s-\mst^2)(\mst^2\,s-\mps^4-\mf^4)\log\left[s\frac{s-2\mps^2-2\mf^2+\mst^2}{\mst^2\,s-(\mps^2-\mf^2)^2}\right]\Bigg\} . \nonumber\\  \label{eq:lamlam}
\eeqn
Computing the reduced cross section from the diagrams at the bottom of Fig.~\ref{Fig:CPTunit} one has to get the same result, with an overall opposite sign. That is indeed what we find, so that Eq.~\eqref{eq:CPT} is satisfied.

We do not report the details of the calculation, but we remark on a few interesting points. The coupling $\lambda_{X_1}$ is purely real, as the fields $S_1$ and $\chi$ are real.
There are two one-loop diagrams to be included for the process $\chi\chi \to U_R \bar\Psi$, commonly referred to as {\em wave} (top - center in the figure) and {\em vertex} (top - right in the figure). Instead, there is only one loop diagram for $\bar U_R \Psi \to U_R \bar\Psi$, which makes this second calculation easier. It is important, however, not to forget the $t$-channel tree-level diagram shown at the bottom - center in the figure. 
The logarithm in Eq.~\eqref{eq:lamlam} comes from the vertical cut on the propagators of $\Psi$ and $\bar U_R$ in the vertex diagram for $\chi\chi \to U_R \bar\Psi$, whereas the same logarithm arises from phase space integration in $\bar U_R \Psi \to U_R \bar\Psi$ and can be traced back to the $t$-channel tree-level diagram.

\bibliographystyle{JHEP}
\bibliography{WIMPy}

\end{document}